\documentclass{IEEEtran}
\usepackage{cite}
\usepackage{amsmath,amssymb,amsfonts}
\usepackage{graphicx}
\usepackage{textcomp}
\usepackage{subcaption}
\usepackage{dblfloatfix}

\def\BibTeX{{\rm B\kern-.05em{\sc i\kern-.025em b}\kern-.08em
    T\kern-.1667em\lower.7ex\hbox{E}\kern-.125emX}}
\begin{document}
\title{A Compact Reconfigurable Antenna for Single-RF-Chain Passive Multi-Target DOA Estimation}
\author{Ruihua Ma, Erik Ralf Algarp, Xiaoping Li, He Huang, Fang Ding, \\ Amir Mohsen Ahmadi Najafabadi, and Anja K. Skrivervik
\thanks{This work was supported by Young Talent Fund of University Association for Science and Technology in Shaanxi, China (No. 20240139), National Natural Science Foundation of China (No. 62371375), Innovation Capability Support Program of Shaanxi (No. 2022TD-37). }
\thanks{Ruihua Ma, Xiaoping Li, and He Huang are with the Key Laboratory  of Equipment Efficiency in Extreme Environment, Ministry of Education, School of Aerospace Science and Technology, Xidian University, Xi'an 710071, China (e-mail: huanghe@xidian.edu.cn). }
\thanks{Ruihua Ma, Erik Ralf Algarp, Fang Ding, Amir Mohsen Ahmadi Najafabadi, and Anja K. Skrivervik are with the Microwaves and Antennas Group, \'Ecole Polytechnique F\'ed\'erale de Lausanne, 1015 Lausanne, Switzerland (e-mail: anja.skrivervik@epfl.ch).}
}

\maketitle

\begin{abstract}
This work proposes a compact and hardware-efficient frequency- and radiation-pattern-reconfigurable antenna (FPRA) for passive multi-target direction-of-arrival (DOA) estimation using a single receive RF chain. The antenna consists of a sectorized circular patch loaded with 16 PIN diodes. By switching the diode states, the current-concentration boundary on the patch is shifted, enabling reconfiguration of both the operating frequency and radiation pattern. With a single receive RF chain, the proposed antenna achieves beam scanning from $-40^\circ$ to $40^\circ$ and provides multiple operating frequencies across the S- and C-bands. Based on these reconfigurable observation states, radiation-pattern switching is used to emulate the spatial sampling of a conventional antenna array, while multi-frequency observations introduce phase diversity to reduce the correlation among echoes from multiple passive targets illuminated by the same transmitter. Experimental results demonstrate that the combined virtual spatial sampling and frequency diversity enable passive multi-target DOA estimation without a conventional antenna array or multiple receive RF chains. The proposed FPRA offers a compact and hardware-efficient sensing solution for future integrated sensing and communication (ISAC) systems.
\end{abstract}


\section{Introduction}
\label{sec:introduction}

\IEEEPARstart{F}{uture} wireless networks are evolving toward integrated sensing, communication, and positioning (ISAC). Besides providing user coverage and data transmission, access-network nodes are also expected to perform environmental sensing and target localization by reusing the same hardware and time--frequency resources~\cite{ref1}. However, the deployment of ISAC nodes is constrained by power consumption, size, cost, and maintenance. Therefore, conventional solutions that improve sensing performance by enlarging the antenna array and increasing the number of radio-frequency (RF) chains are often impractical~\cite{ref2,ref3}. In particular, low-cost and low-power nodes usually adopt a tri-hybrid (digital--analog--antenna) architecture, where the front end is equipped with only one or a few RF chains. This severely limits the observation dimension and makes high-resolution angle estimation and multi-target separation difficult~\cite{ref4,ref5}.

Moreover, future wireless networks are expected to detect and localize passive or non-connected objects, such as chipless RFID tags, pedestrians, vehicles, walls, and other non-cooperative scatterers. When multiple targets are illuminated by the same probing waveform, their echoes are superposed at the receiver and are often highly correlated~\cite{ref6,ref7}. This correlation may make the sample covariance matrix rank deficient, causing classical subspace-based methods, such as MUSIC and ESPRIT, to produce merged or spurious peaks or even fail~\cite{ref8}. The problem is particularly severe in compact single-RF-chain ISAC nodes, which provide only a limited number of observations and lack the spatial diversity needed to decorrelate the echoes and recover the covariance rank. Therefore, obtaining sufficient observation diversity and reducing echo correlation without adding RF chains remains a key challenge for reliable sensing in hardware-constrained receivers~\cite{ref9}. To address this limitation, a natural solution is to introduce an additional reconfigurable electromagnetic aperture between the transmitter, targets, and receiver. Reconfigurable intelligent surfaces (RISs) and programmable metasurfaces can induce state-dependent changes in the phase and amplitude of the incident field or apply space–time coding profiles. Therefore, by switching among different surface coding states, the signal propagation characteristics are intentionally modified, and the receiver can collect multiple independent or quasi-independent measurements without directly increasing the number of RF chains. In this sense, an RIS can act as an external virtual aperture that provides additional spatial and spectral observation diversity.

Based on this principle, programmable electromagnetic surfaces have been investigated for ISAC and sensing enhancement. Ref.~\cite{ref10} proposed an ISAC framework based on a space-time-coding metasurface, where the programmable surface controls the fundamental wave and generates spatially distributed harmonics for sensing. Ref.~\cite{ref11} developed an adaptive programmable metasurface that senses the incident field and updates its reflection matrix in real time for beam control. Ref.~\cite{ref12} introduced a space-time-coding digital metasurface with time-varying reflection coefficients, enabling harmonic beam steering and spectral power redistribution. These studies show that programmable surfaces can provide additional configurable measurement states and therefore help alleviate the observation-dimensionality limitation of RF-chain-constrained sensing systems.

However, external RISs and programmable metasurfaces are not always suitable for compact ISAC access nodes. They usually require large panels and additional installation space to provide sufficient link gain, which increases deployment cost and hardware complexity. Moreover, the optimized surface configuration is highly dependent on the deployment environment and channel calibration~\cite{ref13,ref14}. Once the scenario changes, the surface coding and channel model need to be recalibrated, which motivates the integration of reconfigurability directly into the antenna aperture.
Ref.~\cite{ref15} proposed a boundary-tunable antenna that modifies the cavity modal distribution by switching boundary conditions. This antenna can generate a set of highly distinct radiation patterns for computational imaging, replacing conventional parallel sampling by multiple antenna elements with boundary switching. Ref.~\cite{ref16} studied a frequency-diverse computational imaging antenna and showed that sensing performance can be improved by increasing the spatial diversity of the radiated measurement modes, thereby enhancing the quality of the measurement matrix used for imaging. Moreover, Ref.~\cite{ref17} developed a single-RF-chain direction-finding scheme based on a reconfigurable Alford loop antenna operating at sub-6 GHz, where multiple antenna states are used to create distinct observation patterns for DOA estimation instead of relying on a conventional multi-element array.

Although previous studies have shown that radiation-pattern switching can increase the number of observations with only a few RF chains, future ISAC systems are also expected to support the localization of passive and non-cooperative targets while avoiding deployment-dependent aperture configuration and channel calibration. To address these issues, we propose a frequency and radiation pattern reconfigurable antenna (FPRA). The antenna is based on a circular patch with multiple narrow slots and 16 integrated PIN diodes. With this design, the surface current tends to concentrate near the slot boundaries between sectors connected to ON-state and OFF-state diodes. By switching the diode states, the current-concentration boundary can be shifted over the patch, which enables reconfiguration of the resonant path and radiation pattern. As a result, the antenna can generate a diverse set of beam and frequency states on the same hardware without adding RF chains. Each coding state is treated as an antenna-domain observation state, and the received responses under different states are combined to form an equivalent sensing matrix. In this way, conventional array-based parallel sampling is replaced by coding-based sampling. Moreover, integrating the reconfigurable aperture into the antenna eliminates the need for a separately deployed programmable surface and reduces deployment and calibration complexity. Besides, the antenna provides multi-frequency observations, which introduce different phase shifts for echoes from different targets. This helps reduce echo correlation and improves target separability. Therefore, the proposed FPRA provides a scalable and low-power solution for multi-target localization in future ISAC access nodes with only a single receive RF chain.

\section{Antenna Design and Operation Principle}
\subsection{Antenna structure}

\begin{figure}[!h]
\centering

\begin{subfigure}{0.48\columnwidth}
    \centering
    \includegraphics[width=\linewidth]{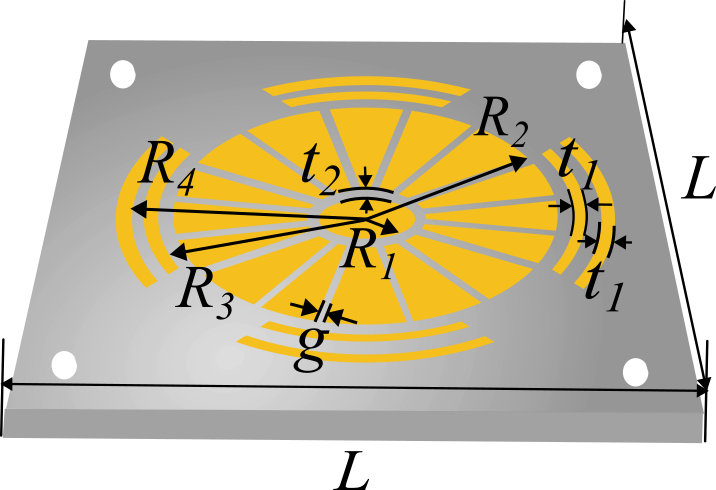}
    \caption{}
\end{subfigure}
\hfill
\begin{subfigure}{0.48\columnwidth}
    \centering
    \includegraphics[width=\linewidth]{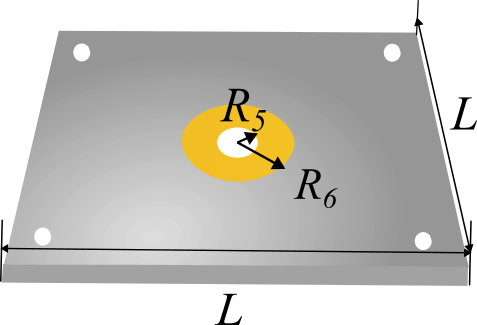}
    \caption{}
\end{subfigure}

\vspace{0.5em}

\begin{subfigure}{0.9\columnwidth}
    \centering
    \includegraphics[width=\linewidth]{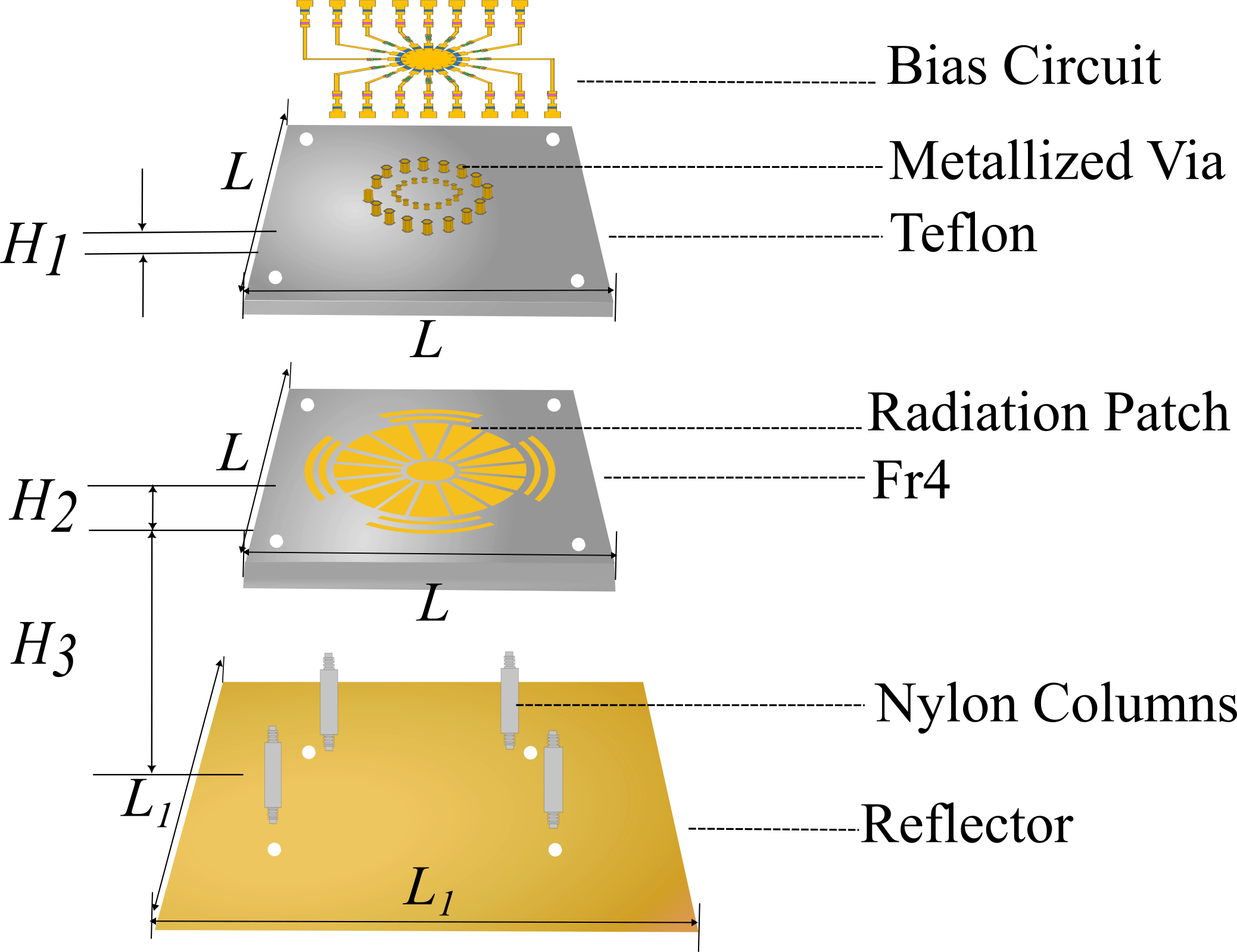}
    \caption{}
\end{subfigure}

\caption{Antenna configuration. (a) Front side of the radiating patch. Dimensions (mm): $L=60$, $R_1=6$, $R_2=19$, $R_3=21$, $R_4=25$, $g=0.2$, $t_1=2$, and $t_2=2$. (b) Back side of the antenna showing the RF ground plane: $R_5=4.6$ and $R_6=12$. (c) 3-D view of the antenna: $L_1=104$, $H_1=1$, $H_2=3$, and $H_3=16$.}

\label{fig:antenna_configuration}
\end{figure}

To implement the proposed localization framework, a frequency- and pattern-reconfigurable antenna (FPRA) is designed, as shown in Fig.~\ref{fig:antenna_configuration}. The antenna uses a $3~\mathrm{mm}$-thick FR-4 substrate. The front side of this substrate, shown in Fig.~\ref{fig:antenna_configuration}(a), carries the radiating patch, while the back side, shown in Fig.~\ref{fig:antenna_configuration}(b), serves as the RF ground plane. The radiating patch consists of a central circular disk and an outer annular patch divided into 16 fan-shaped sectors. Adjacent sectors are separated by narrow gaps of $0.2~\mathrm{mm}$, which interrupt the continuous surface-current path on the patch. Several arc-shaped parasitic branches are arranged outside the annular patch to enhance the radiation directivity. The antenna is fed through a single coaxial RF port, whose inner conductor
is connected to the radiating patch and whose outer conductor is connected
to the RF ground plane.

Fig.~\ref{fig:antenna_configuration}(c) shows the 3-D view of the antenna. A DC bias circuit is implemented on a $1~\mathrm{mm}$-thick Teflon substrate placed above the FR-4 antenna substrate. The 16 PIN diodes (BAR50-02V from Infineon Company) are mounted on the bias-circuit layer. Sixteen metallized vias are used to connect the positive terminals of the PIN diodes to the inner edges of the corresponding outer annular sectors on the radiating patch, while the negative terminals of the PIN diodes are connected to the innermost circular patch through another set of 16 metallized via holes. Therefore, each PIN diode forms a controllable connection branch between the central circular patch and one outer fan-shaped sector. A metallic reflector is placed $16~\mathrm{mm}$ below the FR-4 antenna substrate and is supported by nylon columns. The reflector suppresses backward radiation and improves the front-to-back ratio and gain.

\begin{figure}[!h]
\centerline{\includegraphics[width=\columnwidth]{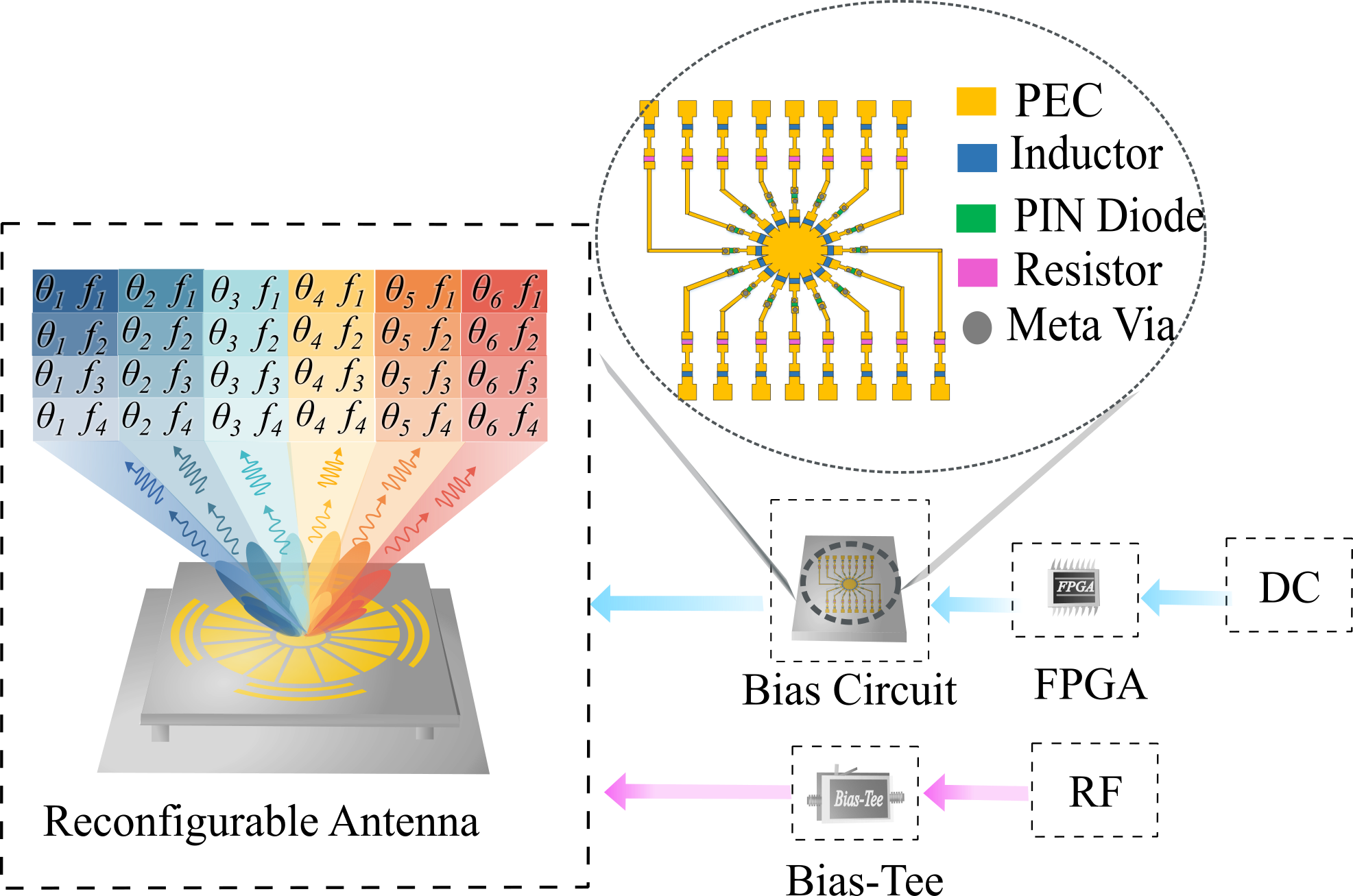}}
\caption{RF and DC control chain of the proposed FPRA, including the bias circuit, FPGA, and DC blocker.}
\label{fig:rf_dc_chain}
\end{figure}

The RF and DC control chain of the proposed FPRA is illustrated in Fig.~\ref{fig:rf_dc_chain}. The RF signal is fed to the antenna through a DC blocker to isolate the RF measurement path from the DC bias. The FPGA applies the control voltages to the 16 PIN diodes through the bias circuit. In each bias branch, a current-limiting resistor protects the PIN diode from excessive current, while an inductor suppresses RF leakage into the DC control network. By changing the FPGA output states, different PIN-diode coding states can be generated for antenna reconfiguration.

\subsection{Current distribution}
\begin{figure}[!h]

\centerline{\includegraphics[width=0.9\columnwidth]{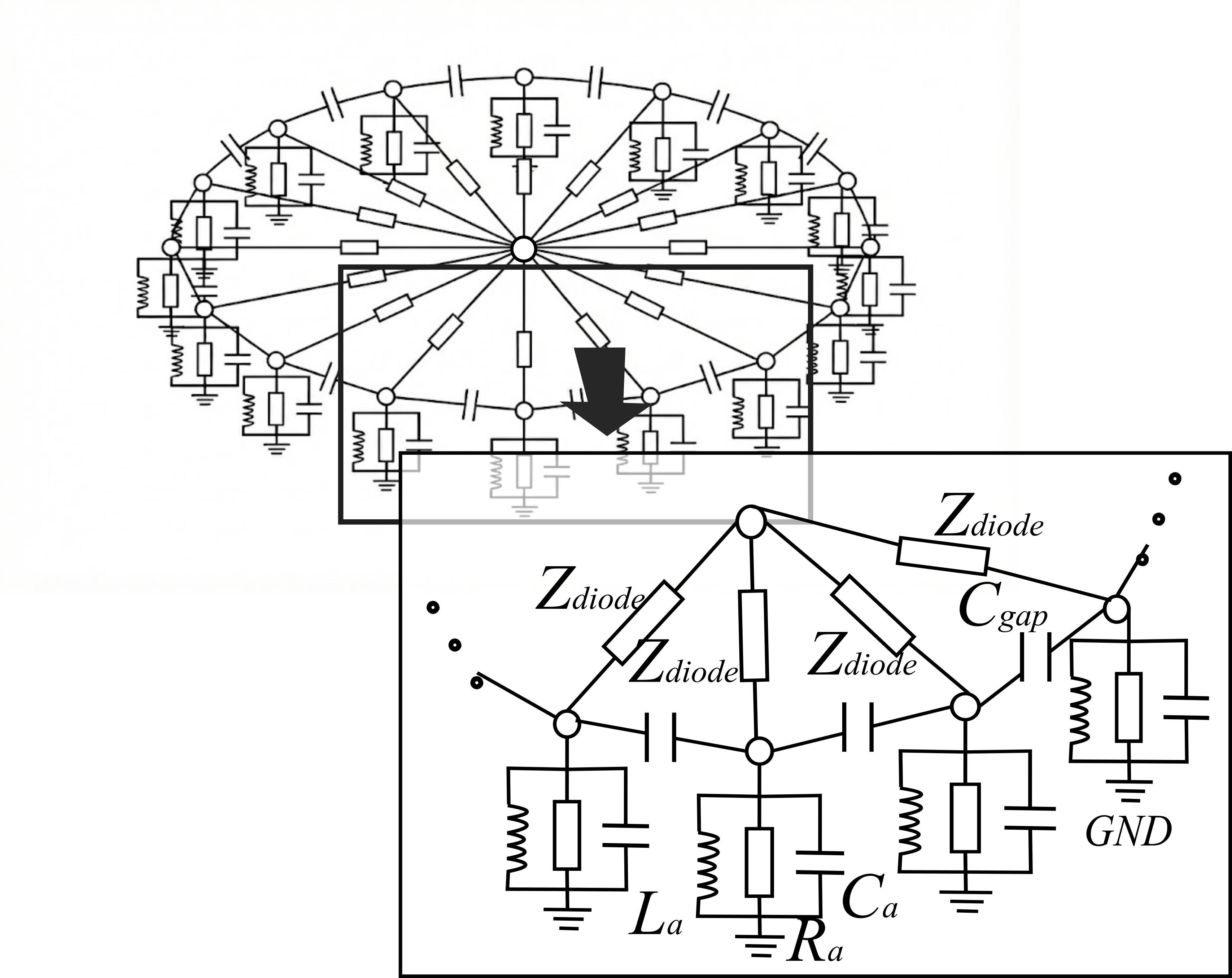}}
\caption{Equivalent circuit of the reconfigurable antenna. $Z_{\mathrm{diode}}$ denotes the equivalent impedance of the PIN diode (ON: $Z_{\mathrm{on}}$, OFF: $Z_{\mathrm{off}}$). $C_{\mathrm{gap}}$ models the capacitive coupling across the $g=0.2$~mm slit between adjacent sector patches. $L_a$, $R_a$, and $C_a$ are the antenna equivalent inductance, resistance, and capacitance, respectively.}
\label{fig:equivalent_circuit}
\end{figure}

\begin{figure}[!h]
\centering
\begin{subfigure}[b]{0.47\columnwidth}
  \centering
  \includegraphics[width=0.7\linewidth]{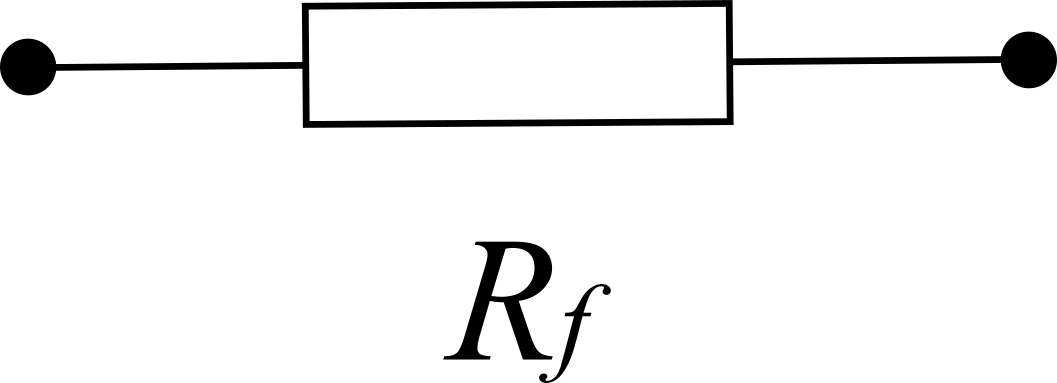}
  \caption{}
  \label{fig:diode_on}
\end{subfigure}
\hfill
\begin{subfigure}[b]{0.47\columnwidth}
  \centering
  \includegraphics[width=0.7\linewidth]{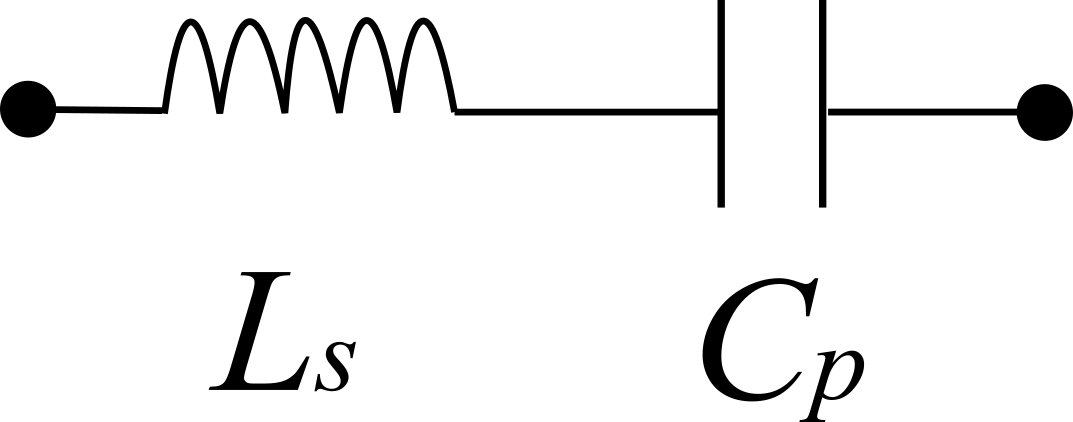}
  \caption{}
  \label{fig:diode_off}
\end{subfigure}

\caption{Equivalent circuit model of the PIN diode at microwave frequencies.
(a) On state. (b) OFF state. $L_s = 0.6~\mathrm{nH}$, $R_f = 1.2~\Omega$, $C_p = 0.15~\mathrm{pF}$.}
\label{fig:diode_model}
\end{figure}

In the equivalent circuit of the proposed antenna shown in Fig.~\ref{fig:equivalent_circuit}, each sector is connected to the central patch through a switchable branch denoted by $Z_{\mathrm{diode}}$, while the slits between adjacent sectors are modeled by the coupling capacitor $C_{\mathrm{gap}}$. The impedance of $Z_{\mathrm{diode}}$ is determined by the equivalent circuit of the PIN diode shown in Fig.~\ref{fig:diode_model}~\cite{ref18}.

For a sector connected to an ON-state diode, $Z_{\mathrm{on}}$ is modeled by the series resistance $R_f$, which provides a low-impedance electrical connection between the sector and the feed node. Therefore, the RF boundary condition of this sector is strongly constrained by the excitation source.

For a sector connected to an OFF-state diode, $Z_{\mathrm{off}}$ is modeled by the series inductance $L_s$ and the junction capacitance $C_p$~\cite{ref18}, so only weak capacitive coupling exists between the sector and the feed. Therefore, this sector is mainly excited by mutual coupling from adjacent sectors through $C_{\mathrm{gap}}$.

Accordingly, the sectors connected to ON-state diodes are regarded as actively driven radiating regions, whereas the sectors connected to OFF-state diodes behave as passively coupled parasitic regions. Therefore, the voltage magnitude and phase near the boundary slit between an ON-state sector and an OFF-state sector differ significantly.

This boundary slit is also a metal edge of the patch. Due to edge-field enhancement, the electric-field lines become denser in the slit region. In the dielectric inside the slit, the total current density is dominated by the displacement-current density,
\begin{equation}
\mathbf{J}\approx\mathbf{J}_d=j\omega\varepsilon_g\mathbf{E},
\end{equation}
where $\varepsilon_g$ is the permittivity of the slit medium, $\mathbf{E}$ is the electric field across the slit, and $\mathbf{J}_d$ denotes the displacement-current density. From the integral form of the Amp\`ere--Maxwell law,
\begin{equation}
\oint_C \mathbf{H}\cdot d\mathbf{l}
\approx \iint_S \mathbf{J}_d\cdot d\mathbf{S},
\end{equation}
a stronger electric field across the slit leads to a larger displacement current, which in turn enhances the magnetic field in the vicinity of the slit.

Furthermore, on a perfect electric conductor (PEC) surface, the surface-current density is related to the discontinuity of the magnetic field across the conductor surface by
\begin{equation}
\mathbf{J}_s=\hat{\mathbf{n}}\times(\mathbf{H}_{\mathrm{out}}-\mathbf{H}_{\mathrm{in}}),
\end{equation}
where $\hat{\mathbf{n}}$ is the outward unit normal vector of the metal surface, $\mathbf{H}_{\mathrm{out}}$ is the magnetic field just outside the conductor surface, and $\mathbf{H}_{\mathrm{in}}$ is the magnetic field inside the conductor. For a PEC, the internal field is approximately zero, i.e., $\mathbf{H}_{\mathrm{in}}\approx 0$, and thus
\begin{equation}
\mathbf{J}_s \approx \hat{\mathbf{n}}\times \mathbf{H}_{t},
\end{equation}
where $\mathbf{H}_t$ denotes the tangential magnetic field just outside the metal surface. Therefore, enhancement of the tangential magnetic field near the slit edge directly leads to an increase in the local surface-current magnitude $|\mathbf{J}_s|$, resulting in the current-concentration effect shown in Fig.~\ref{fig:current_distribution}.

When two adjacent sectors are both connected to ON-state diodes, their equivalent impedances to the central patch are both $Z_{\mathrm{on}}$. Since the two sectors are constrained by the same feed and have identical geometry, their electrical responses tend to be similar. As a result, the electric field and displacement current across the slit are weak, and the two sectors can be approximately regarded as a single connected patch. Therefore, no obvious current-concentration effect appears at the slit in the simulated current distribution.

\subsection{Frequency Reconfigurability}

The frequency reconfigurability is mainly attributed to the change in the effective current path on the circular patch. 
\begin{figure}[!h]
\centerline{\includegraphics[width=\columnwidth]{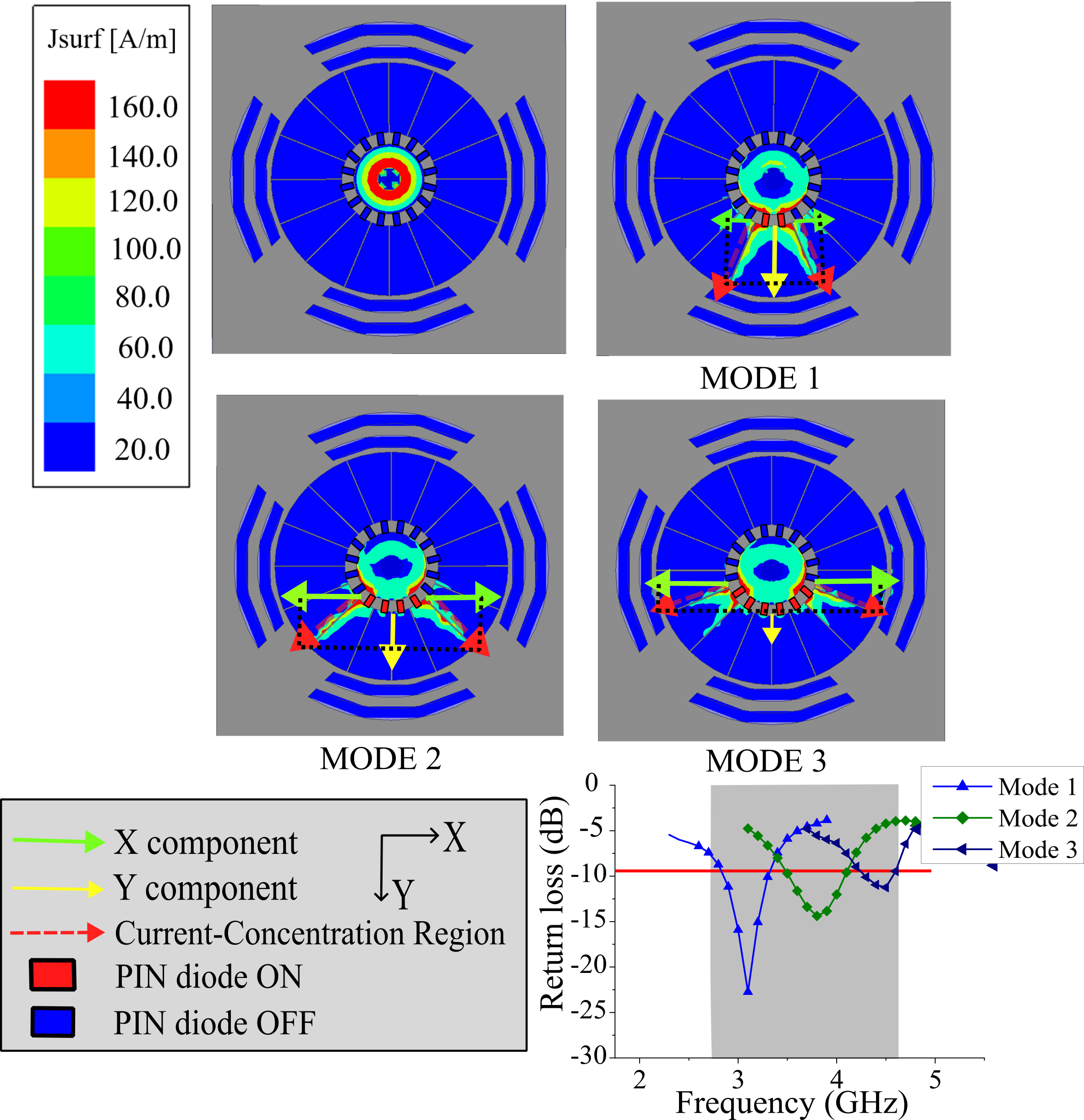}}
\caption{Simulated surface-current distributions of the proposed reconfigurable antenna under different PIN-diode coding states: Mode~1 at 3.1~GHz, Mode~2 at 3.6~GHz, and Mode~3 at 4.4~GHz.}
\label{fig:current_distribution}
\end{figure}
As shown in Fig.~\ref{fig:current_distribution}, the PIN-diode coding changes the position and orientation of the current-concentration boundary between the ON-diode-connected and OFF-diode-connected regions. In the figure, the red rectangles denote ON-state diodes, whereas the blue rectangles denote OFF-state diodes. When only a few sectors near the $y$-axis are connected, the dominant current path is mainly along the $y$-direction, resulting in a relatively long effective resonant length.

As more sectors are connected toward the $x$-direction, the conductive region expands laterally and the current-concentration boundary is shifted accordingly. Due to the symmetry with respect to the $y$-axis, the transverse current components on the left and right sides largely cancel each other, while the effective current contributing to the main resonance is mainly along the $y$-direction. Therefore, the effective resonant path length, denoted by $L_{\mathrm{eff}}$, is shortened, leading to an upward shift of the resonant frequency.

Therefore, by controlling the PIN-diode coding state, the dominant current path and the corresponding $L_{\mathrm{eff}}$ can be adjusted. This enables frequency reconfiguration without changing the physical size of the antenna.

\subsection{Radiation pattern Reconfigurability}

\begin{figure}[!h]

\centerline{\includegraphics[width=\columnwidth]{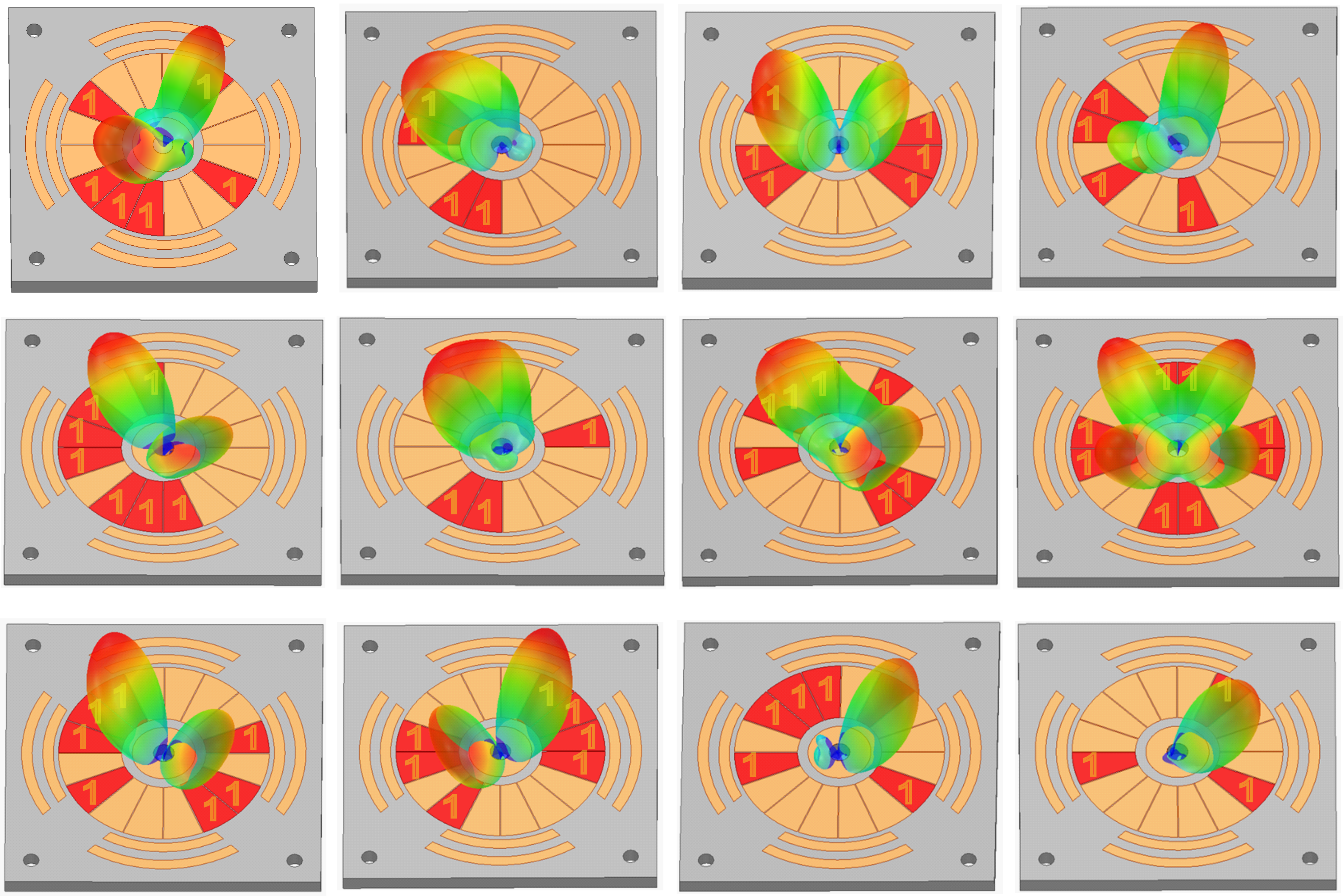}}
\caption{Radiation patterns of the proposed antenna under different PIN-diode coding states.}
\label{fig:radiation_patterns}
\end{figure}

The radiation-pattern reconfiguration is produced by the same PIN-diode coding mechanism. When the diode states are changed, the sectors connected to ON-state diodes form the main driven region, while the sectors connected to OFF-state diodes are excited mainly through mutual coupling and behave as parasitic regions. As a result, switching the diode states effectively reconfigures the phase distribution over the entire aperture. This creates an ``active and parasitic'' phase profile, which enables beam steering as shown in Fig.~\ref{fig:radiation_patterns}. This behavior can be interpreted from the active-element-pattern perspective, where the radiation of each driven sector is modified by adjacent parasitic sectors and mutual coupling~\cite{ref19,ref20}.
Moreover, because the sectors are uniformly arranged, many diode-coding states are geometrically equivalent under cyclic rotations and mirror symmetry. Therefore, a structured beam codebook can be built by selecting a small set of reference codes and generating additional beam codeword entries via circular shifts of the sector-wise diode states.

\subsection{Genetic Algorithm}
With 16 PIN diodes providing a large number of degrees of freedom, the antenna can generate a rich set of radiation modes. To ensure these patterns achieve optimal performance and comprehensive coverage, this work introduces an intelligent search into the reconfigurable antenna to identify the best beam codebook for DOA estimation applications.

Accordingly, the primary task of the beam codebook design is to select a set of PIN diode configurations that simultaneously satisfy prescribed radiation specifications--main-beam direction, half-power beamwidth (HPBW), sidelobe level (SLL), and peak gain and operating constraints, including $|S_{11}|$ and efficiency. Furthermore, the algorithm ensures that the patterns across different states are sufficiently diverse and span a wide steering range, providing complementary spatial responses for subsequent DOA estimation observations.
\begin{figure}[!h]

\centerline{\includegraphics[width=\columnwidth]{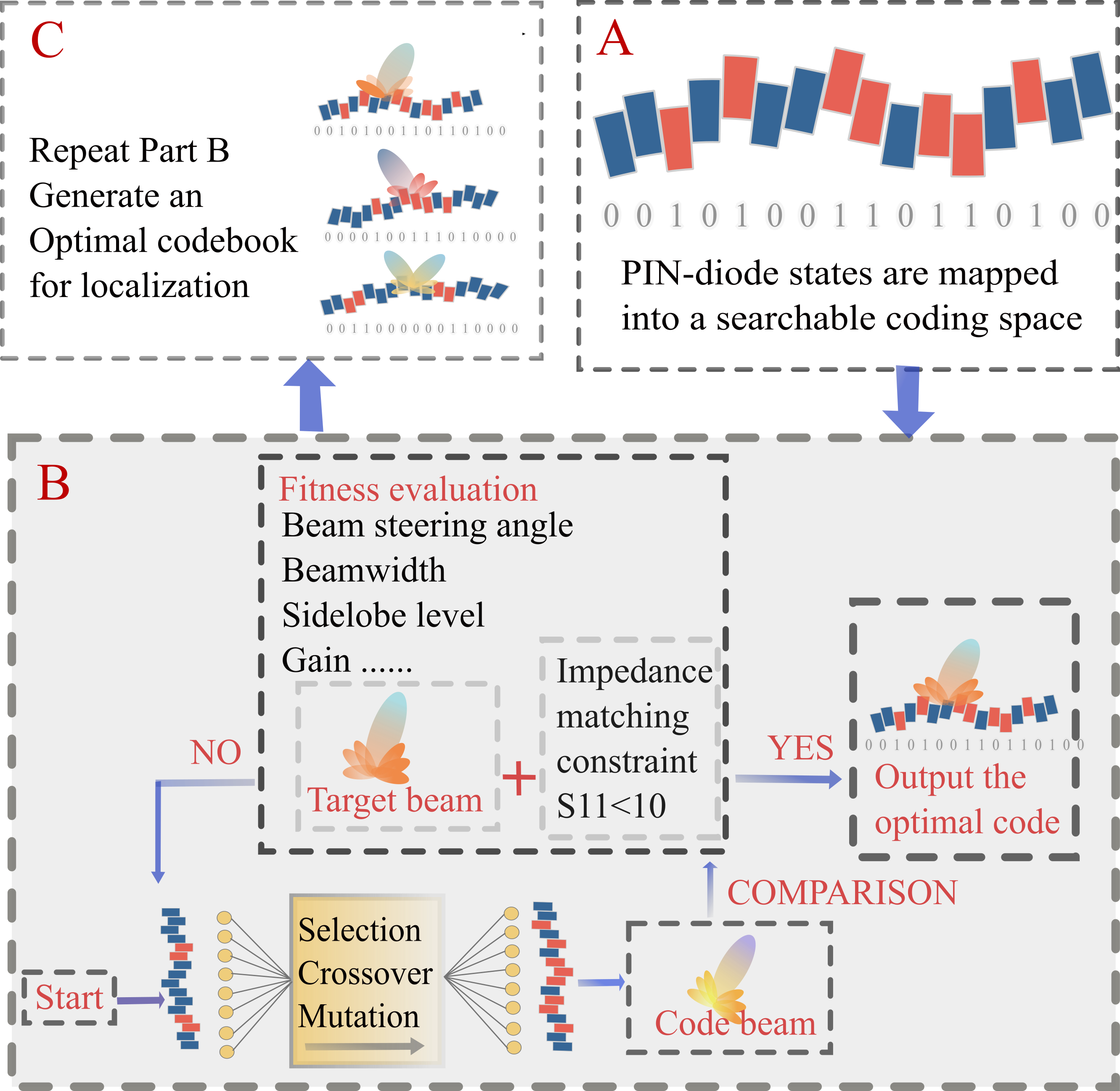}}
\caption{Flowchart of the genetic algorithm based beam-codebook optimization procedure.}

\label{fig:ga_flowchart}
\end{figure}

The genetic algorithm naturally utilizes a binary string as the decision variable, where each bit corresponds to a "gene." We represent each PIN-state combination as a length-16 binary vector:
\begin{equation}
\mathbf{b} = [b_1, b_2, \ldots, b_{16}]^{T}
\end{equation}
The optimization takes place within this coding space, where each bit indicates the ON/OFF state of a specific PIN diode. The fitness function is designed to balance beam performance: gain and beamwidth, with essential requirements: return loss and efficiency. By simulating biological evolution through genetic operations, the algorithm identifies a diverse set of switching configurations that meet requirements while delivering superior radiation patterns. The flowchart of the genetic algorithm is shown in Fig.~\ref{fig:ga_flowchart}.

\begin{figure}[!t]
\centering

\begin{subfigure}{0.7\columnwidth}
    \centering
    \includegraphics[width=\linewidth]{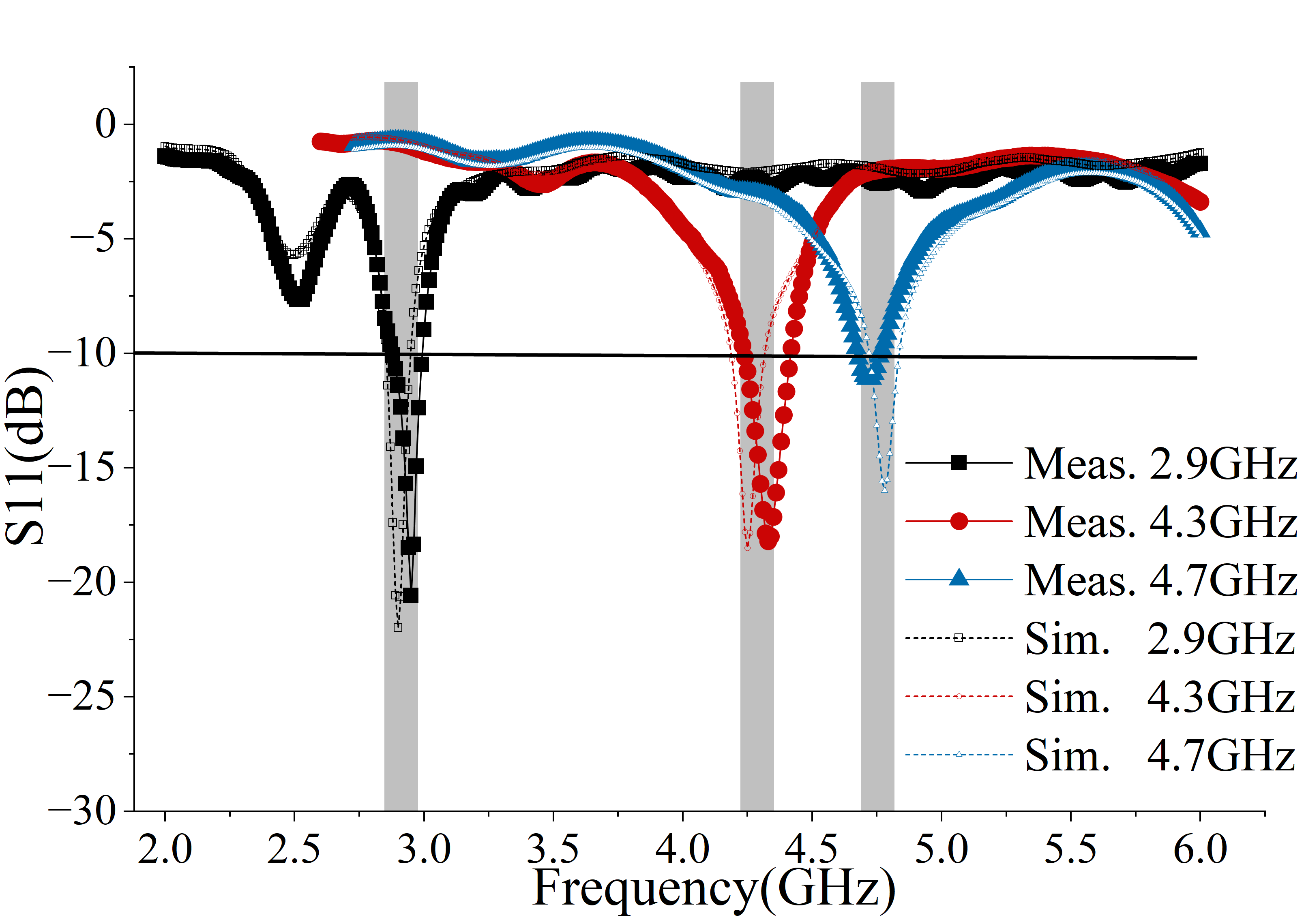}
    \caption{}
\end{subfigure}
\hfill
\begin{subfigure}{0.7\columnwidth}
    \centering
    \includegraphics[width=\linewidth]{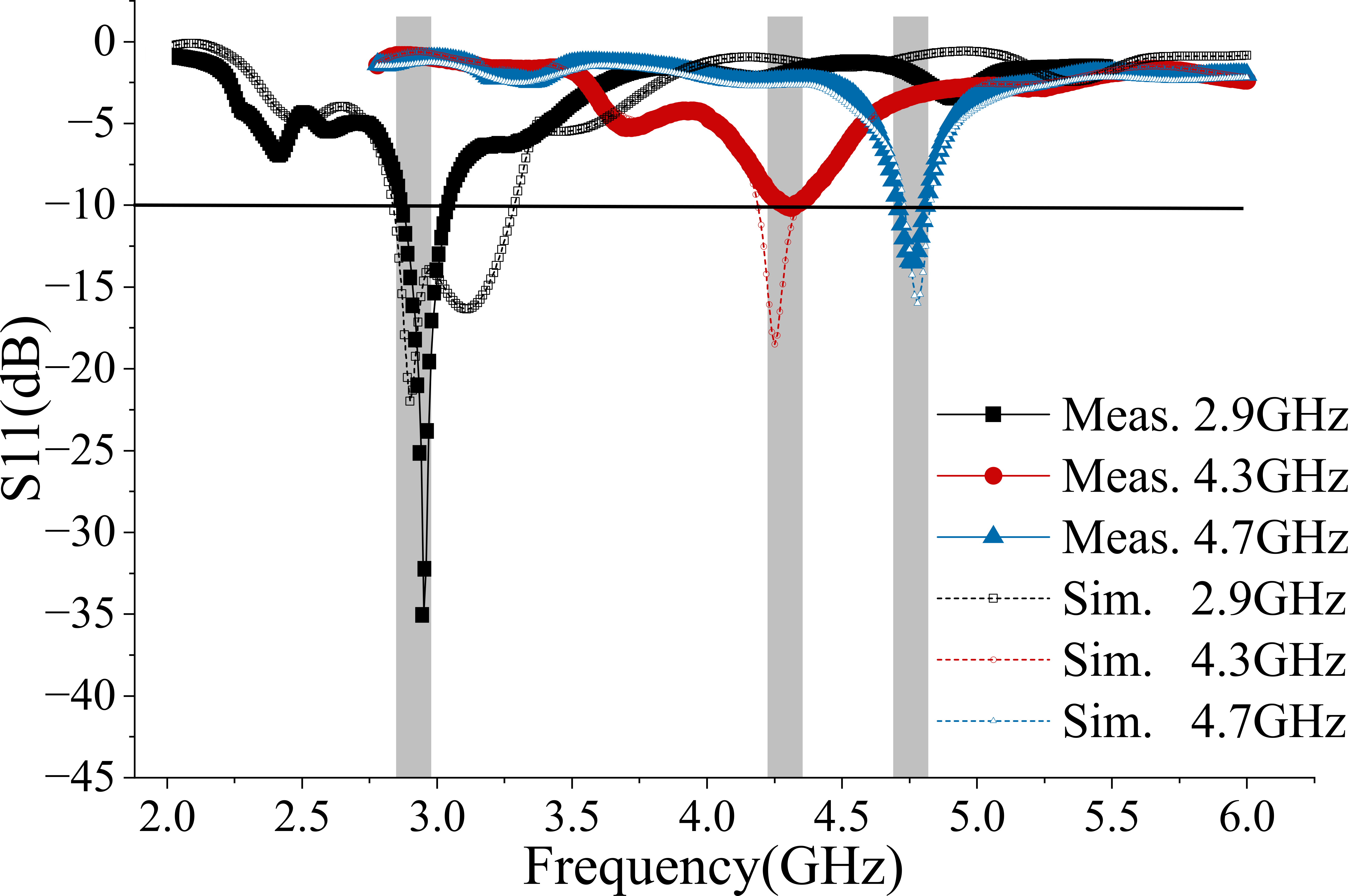}
    \caption{}
\end{subfigure}

\vspace{0.5em}

\begin{subfigure}{0.7\columnwidth}
    \centering
    \includegraphics[width=\linewidth]{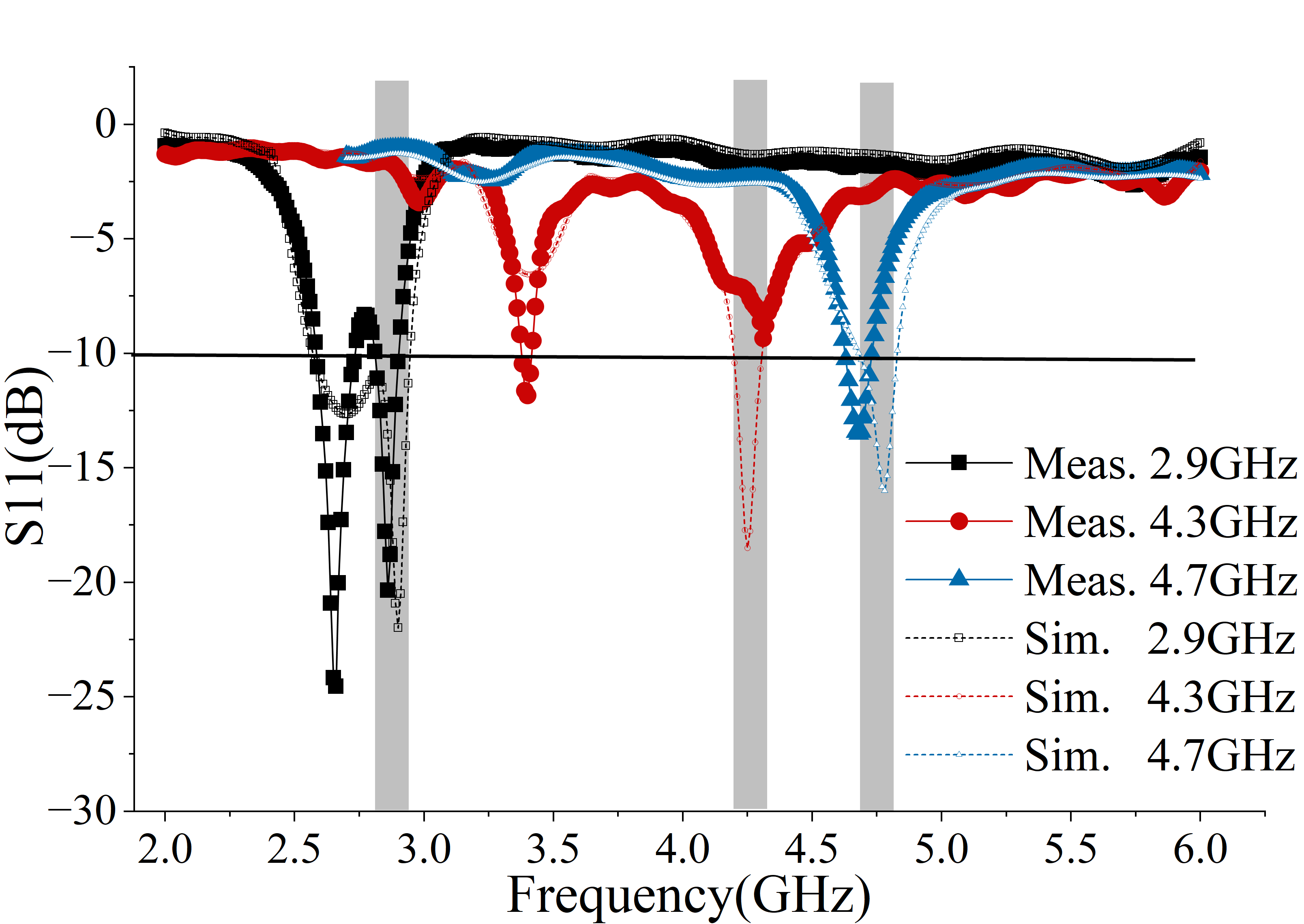}
    \caption{}
\end{subfigure}
\hfill
\begin{subfigure}{0.7\columnwidth}
    \centering
    \includegraphics[width=\linewidth]{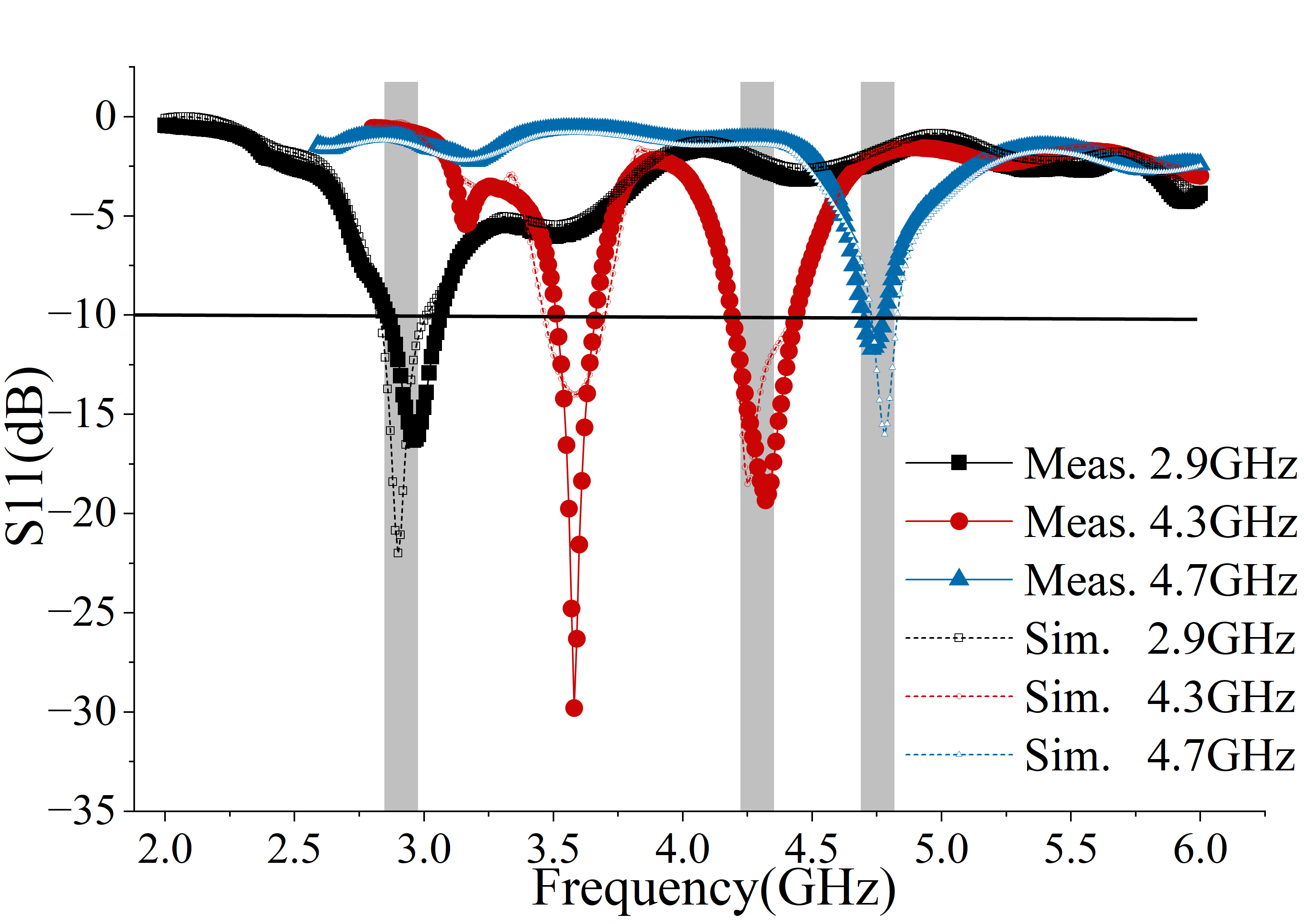}
    \caption{}
\end{subfigure}

\caption{Simulated and Measured reflection coefficients of the antenna for different beam states at the selected frequencies: (a) Beam 1, (b) Beam 2, (c) Beam 3, and (d) Beam 4.}
\label{fig:s11_results}
\end{figure}

\begin{figure}[!h]
\centering

\begin{subfigure}{0.65\columnwidth}
    \centering
    \includegraphics[width=\linewidth]{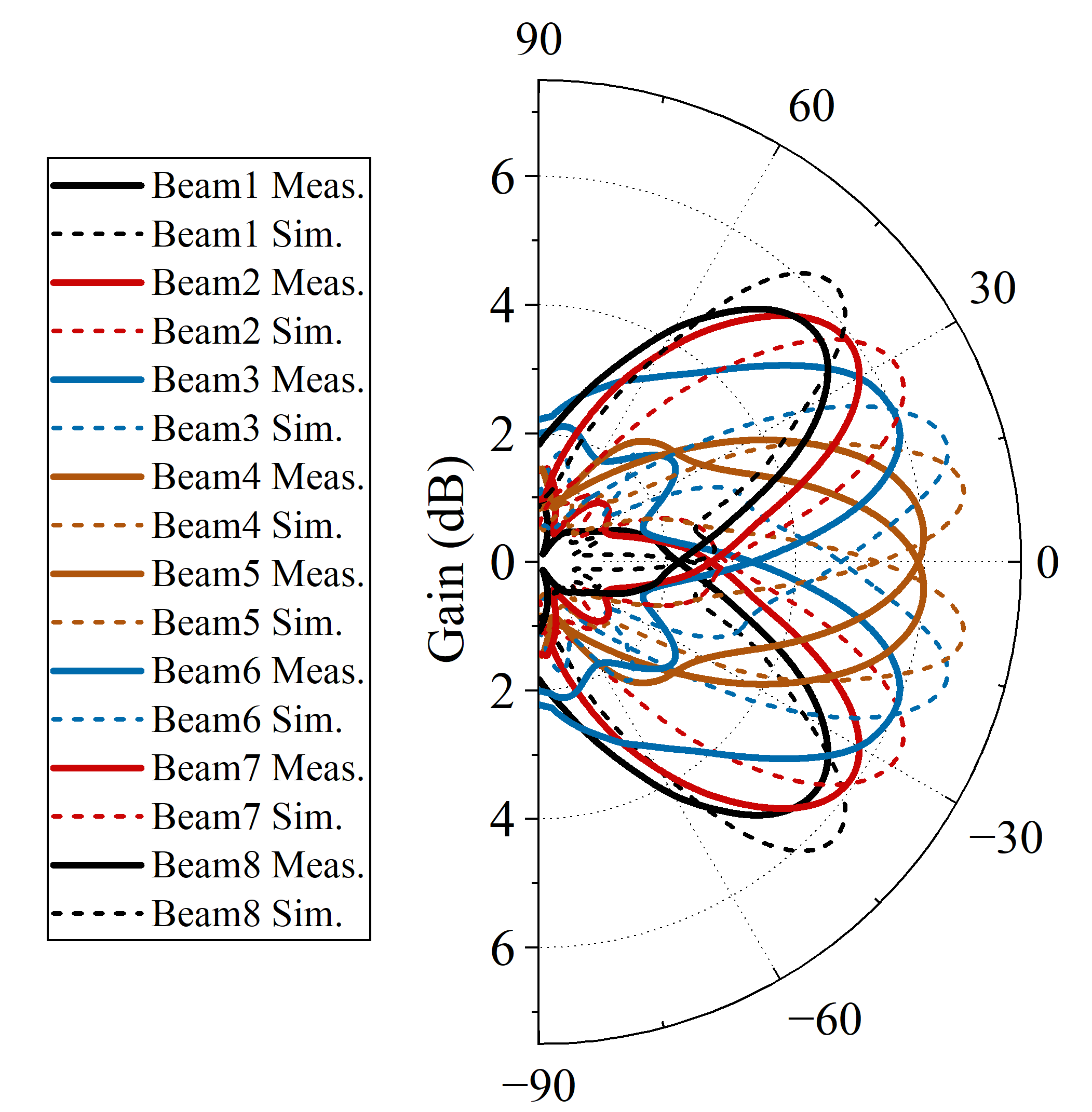}
    \caption{}
\end{subfigure}

\vspace{0.5em}

\begin{subfigure}{0.45\columnwidth}
    \centering
    \includegraphics[width=\linewidth]{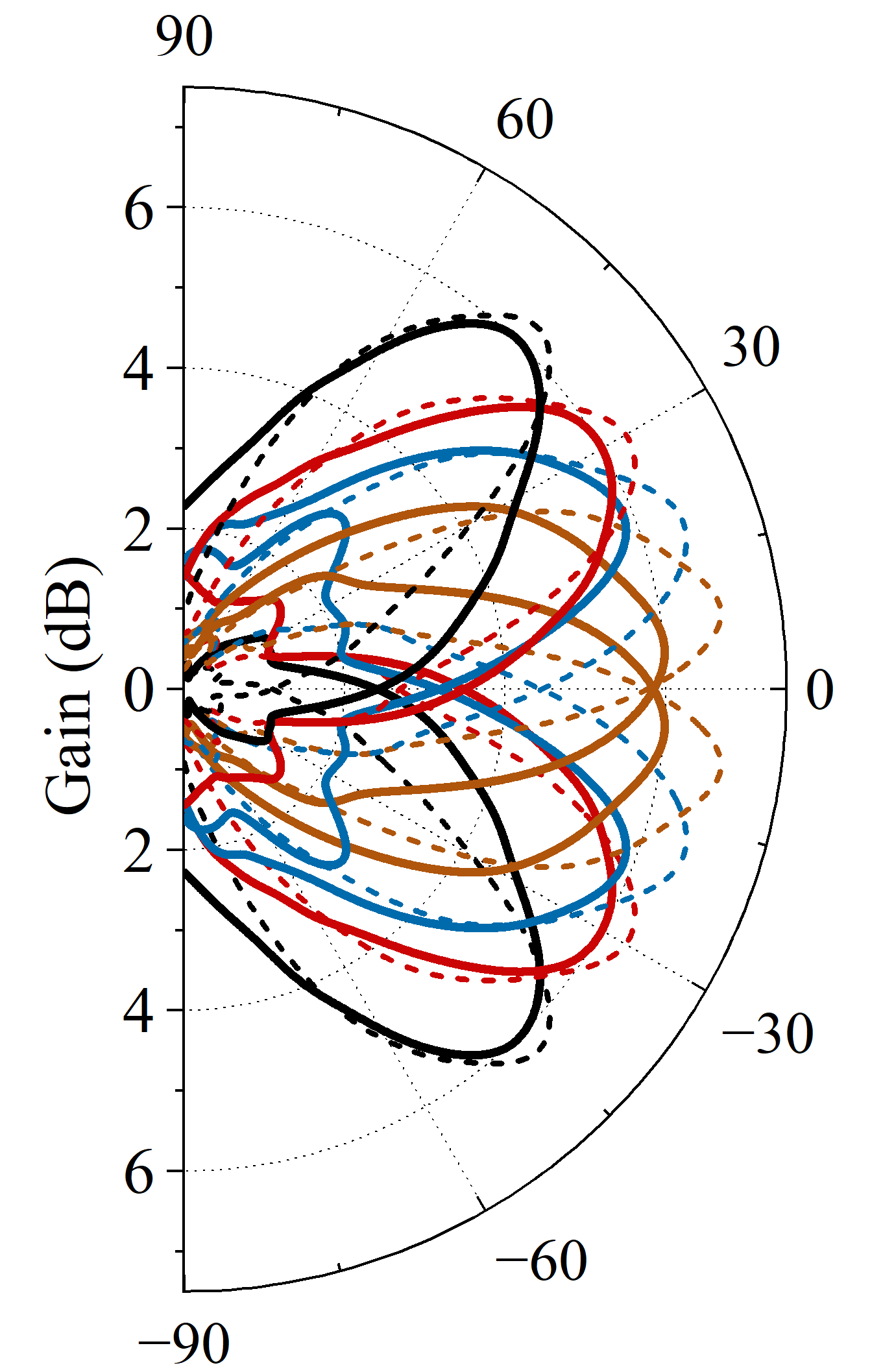}
    \caption{}
\end{subfigure}
\hfill
\begin{subfigure}{0.45\columnwidth}
    \centering
    \includegraphics[width=\linewidth]{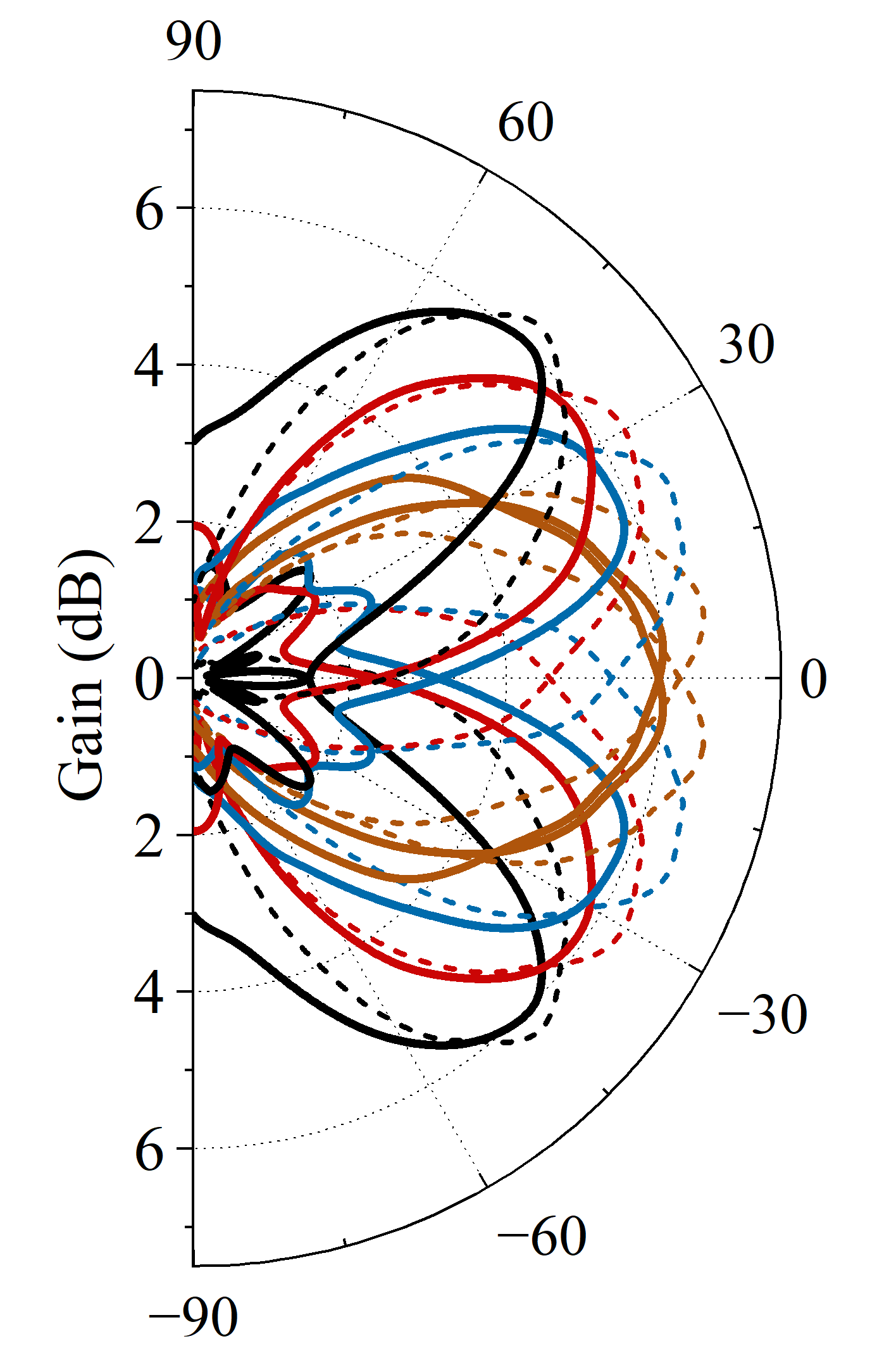}
    \caption{}
\end{subfigure}

\caption{Measured and simulated radiation patterns of the proposed antenna at three operating frequencies: (a) 2.9 GHz, (b) 4.3 GHz, and (c) 4.7 GHz.}
\label{fig:radiation_measurement}
\end{figure}

To ensure accuracy, the evaluation process is automated using a Python-HFSS co-simulation loop. The script automatically updates the antenna model with new PIN-diode configurations, executes the electromagnetic simulations in HFSS, and feeds the simulated results back into the GA to refine the next generation. This automated cycle continues until a set of switching states satisfies the predefined design criteria. The $S_{11}$ responses and radiation patterns corresponding to the GA-selected coding states are presented in Fig.~\ref{fig:s11_results} and shown in Fig.~\ref{fig:radiation_measurement}, respectively. These selected states are subsequently used to construct the ``antenna-domain measurement operator'' for DOA estimation.

\section{DOA Estimation Algorithm}
\label{sec:doa_algorithm}

\begin{figure*}[!t]
\centering
\includegraphics[width=\textwidth]{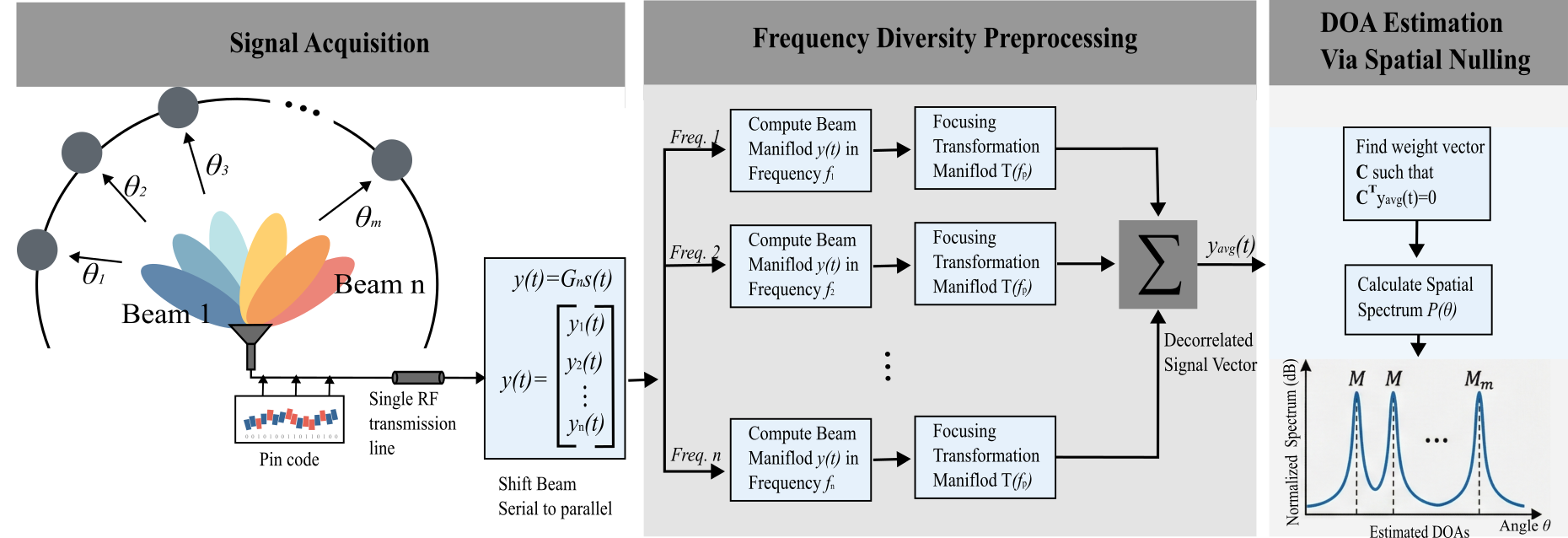}
\caption{Flowchart of the proposed DOA algorithm.}
\label{fig:algorithm}
\end{figure*}

In a conventional antenna-array-based MUSIC algorithm, the direction information is encoded by the phase differences among spatially separated antenna elements. Therefore, the array steering vector describes how a plane wave arriving from a given direction is observed by different physical elements.

In the proposed single-RF-chain system, multiple radiation-pattern responses are obtained by switching the antenna coding states. These different patterns observe the same incident wave with different angle-dependent complex weights. Therefore, the direction information is encoded not by the spatial phase differences among physical elements, but by the response differences among the multiple radiation patterns generated by different coding states. In this sense, the coding states provide virtual observation channels, and their calibrated pattern responses form a code-state manifold analogous to the conventional array manifold.

The radar received signal from one target can be written as
\begin{equation}
y(t)
=
\sqrt{
\frac{
G_T(\theta_t)G_R(\theta_r)\lambda^2\delta
}{
(4\pi)^3R_t^2R_r^2
}
}
x(t-\tau),
\end{equation}
where $y(t)$ is the received signal, $x(t)$ is the transmitted
signal, and $t$ denotes time. $G_T(\theta_t)$ and
$G_R(\theta_r)$ denote the gains of the transmit antenna and the
proposed reconfigurable receive antenna in the target direction,
respectively. $\theta_t$ and $\theta_r$ are the target angles
relative to the transmit and receive antennas, respectively.
$R_t$ and $R_r$ are the distances from the transmit antenna to
the target and from the target to the receive antenna,
respectively. Moreover, $\lambda$ is the wavelength, $\delta$ is
the radar cross section (RCS), and $\tau$ is the propagation
delay. In the following pattern-domain signal model, the
receive-side angular response is absorbed into the calibrated
complex response $\Phi_i(\theta,f)$.

For $n$ reconfigurable radiation patterns and $m$ targets, the
received signal vector at frequency $f$ is
\begin{equation}
\mathbf{y}(t,f)
=
\mathbf{G}(f)\mathbf{s}(t,f)
+
\mathbf{w}(t,f),
\end{equation}
where $\mathbf{w}(t,f)$ is the noise vector, and
$\mathbf{y}(t,f)$ is the received signal measured under
different reconfigurable radiation patterns:
\begin{equation}
\mathbf{y}(t,f)
=
\begin{bmatrix}
y_1(t,f)\\
y_2(t,f)\\
\vdots\\
y_n(t,f)
\end{bmatrix}.
\end{equation}

The vector $\mathbf{s}(t,f)$ collects the contributions from the
$m$ targets:
\begin{equation}
\mathbf{s}(t,f)
=
\begin{bmatrix}
s_1(t,f)\\
s_2(t,f)\\
\vdots\\
s_m(t,f)
\end{bmatrix}.
\end{equation}

The matrix $\mathbf{G}(f)\in\mathbb{C}^{n\times m}$ describes
the directional responses of the $n$ radiation patterns toward
the $m$ target directions and is defined as
\begin{equation}
\mathbf{G}(f)
=
\begin{bmatrix}
\Phi_1(\theta_1,f) & \cdots & \Phi_1(\theta_m,f)\\
\vdots & \ddots & \vdots\\
\Phi_n(\theta_1,f) & \cdots & \Phi_n(\theta_m,f)
\end{bmatrix},
\end{equation}
where $\Phi_i(\theta,f)$ denotes the calibrated complex response
of the $i$-th reconfigurable radiation pattern toward direction
$\theta$ at frequency $f$. Therefore, the steering vector
corresponding to direction $\theta$ is
\begin{equation}
\mathbf{g}(\theta,f)
=
\begin{bmatrix}
\Phi_1(\theta,f)\\
\Phi_2(\theta,f)\\
\vdots\\
\Phi_n(\theta,f)
\end{bmatrix}.
\end{equation}

In this formulation, each reconfigurable radiation pattern
provides one distinct observation of the same target echo. Thus,
although the receiver has only one RF chain, switching among
different radiation-pattern states forms a virtual array manifold
in the pattern domain.

To estimate the target DOAs, the covariance matrix of the
received signal is first calculated as
\begin{equation}
\mathbf{R}_y(f)
=
E\left\{
\mathbf{y}(t,f)\mathbf{y}^{H}(t,f)
\right\}.
\end{equation}
In practice, it is estimated from $L$ snapshots as
\begin{equation}
\widehat{\mathbf{R}}_y(f)
=
\frac{1}{L}
\sum_{\ell=1}^{L}
\mathbf{y}(t_\ell,f)
\mathbf{y}^{H}(t_\ell,f).
\end{equation}

The eigenvalue decomposition of $\widehat{\mathbf{R}}_y(f)$ is
given by
\begin{equation}
\widehat{\mathbf{R}}_y(f)
=
\mathbf{E}_s\mathbf{\Lambda}_s\mathbf{E}_s^H
+
\mathbf{E}_n\mathbf{\Lambda}_n\mathbf{E}_n^H,
\end{equation}
where $\mathbf{E}_s$ is the signal subspace spanned by the
dominant eigenvectors, and $\mathbf{E}_n$ is the noise subspace.
According to the MUSIC principle, the steering vectors
corresponding to the true target directions are approximately
orthogonal to the noise subspace:
\begin{equation}
\mathbf{E}_n^H
\mathbf{g}(\theta_k,f)
\approx
\mathbf{0},
\quad
k=1,\ldots,m.
\end{equation}

Therefore, the MUSIC spatial spectrum is
constructed as
\begin{equation}
P_{\mathrm{MUSIC}}(\theta,f)
=
\frac{1}{
\left\|
\mathbf{E}_n^H\mathbf{g}(\theta,f)
\right\|_2^2
}.
\end{equation}
The target DOAs are estimated from the $m$ dominant local maxima
of $P_{\mathrm{MUSIC}}(\theta,f)$.

For multiple passive targets, the reflected signals are often
highly correlated because they are generated by the same
transmitted waveform. At the $p$-th frequency, the echo from the
$k$-th target can be written as
\begin{equation}
s_k(t,f_p)
=
\alpha_k(f_p)
e^{-j2\pi f_p\tau_k}
x(t),
\end{equation}
where $\alpha_k(f_p)$ represents the frequency-dependent complex
propagation and scattering coefficient of the $k$-th target, and
$\tau_k$ is the corresponding propagation delay. Thus, different
target echoes contain the same waveform $x(t)$ but have different
frequency-dependent amplitudes and phases. At a single frequency,
these echoes may remain highly correlated, which can make the
source covariance matrix rank deficient and prevent accurate
separation of the signal and noise subspaces. As a result, the
MUSIC spectrum may become unstable or fail to resolve multiple
targets.

To alleviate this issue, the frequency-reconfigurable
characteristic of the proposed antenna is further utilized. By
changing the PIN-diode coding state, the effective electrical
length of the antenna is modified, thereby providing multiple
frequency observation points. For any target pair with
propagation delays $\tau_i$ and $\tau_j$, the phase-related
correlation factor over $P$ frequencies is
\begin{equation}
\gamma_{ij}
=
\frac{1}{P}
\sum_{p=1}^{P}
e^{-j2\pi f_p(\tau_i-\tau_j)}.
\end{equation}
This term represents the average relative phase difference
between the two target echoes over the selected frequencies.
When the relative phases
$2\pi f_p(\tau_i-\tau_j)$ are sufficiently dispersed, coherent
accumulation is reduced and $|\gamma_{ij}|$ generally becomes
smaller than one. Therefore, appropriately selected frequencies
can reduce the effective correlation between different target
echoes.

Because the calibrated pattern-domain manifold varies with
frequency, covariance matrices obtained at different frequencies
cannot be directly averaged. Therefore, the manifolds at the
selected frequencies are aligned with that at a reference
frequency $f_0$ before covariance averaging.

Let $\mathbf{A}(f_p)$ denote the steering matrix formed by the
calibrated pattern-domain steering vectors over the angular
search grid at frequency $f_p$. A unitary focusing matrix
$\mathbf{T}(f_p)$ is designed to minimize
\begin{equation}
\left\|
\mathbf{T}(f_p)\mathbf{A}(f_p)
-
\mathbf{A}(f_0)
\right\|_F^2,
\end{equation}
subject to
\begin{equation}
\mathbf{T}^{H}(f_p)\mathbf{T}(f_p)
=
\mathbf{I}.
\end{equation}

By applying singular value decomposition,
\begin{equation}
\mathbf{A}(f_0)\mathbf{A}^{H}(f_p)
=
\mathbf{U}_p
\mathbf{\Sigma}_p
\mathbf{V}_p^{H},
\end{equation}
the corresponding focusing matrix is obtained as
\begin{equation}
\mathbf{T}(f_p)
=
\mathbf{U}_p\mathbf{V}_p^{H}.
\end{equation}
For the reference frequency, $\mathbf{T}(f_0)=\mathbf{I}$.
Since $\mathbf{T}(f_p)$ is unitary, the white-noise property is
preserved after focusing.

The covariance matrix at the $p$-th frequency is estimated from
$L$ snapshots as
\begin{equation}
\widehat{\mathbf{R}}_y(f_p)
=
\frac{1}{L}
\sum_{\ell=1}^{L}
\mathbf{y}(t_\ell,f_p)
\mathbf{y}^{H}(t_\ell,f_p).
\end{equation}

The focused covariance matrices are then averaged over the
$P$ selected frequencies:
\begin{equation}
\widehat{\mathbf{R}}_{\mathrm{avg}}
=
\frac{1}{P}
\sum_{p=1}^{P}
\mathbf{T}(f_p)
\widehat{\mathbf{R}}_y(f_p)
\mathbf{T}^{H}(f_p).
\end{equation}

The eigenvalue decomposition of
$\widehat{\mathbf{R}}_{\mathrm{avg}}$ is given by
\begin{equation}
\widehat{\mathbf{R}}_{\mathrm{avg}}
=
\mathbf{E}_s\mathbf{\Lambda}_s\mathbf{E}_s^H
+
\mathbf{E}_n\mathbf{\Lambda}_n\mathbf{E}_n^H,
\end{equation}
where $\mathbf{E}_s$ and $\mathbf{E}_n$ denote the signal and
noise subspaces, respectively.

After frequency focusing, the steering vectors associated with
the true target directions are approximately orthogonal to the
noise subspace:
\begin{equation}
\mathbf{E}_n^H
\mathbf{g}(\theta_k,f_0)
\approx
\mathbf{0},
\quad
k=1,\ldots,m.
\end{equation}

Therefore, the final MUSIC spatial spectrum is calculated at the
reference frequency as
\begin{equation}
P_{\mathrm{MUSIC}}(\theta)
=
\frac{1}{
\left\|
\mathbf{E}_n^H
\mathbf{g}(\theta,f_0)
\right\|_2^2
}.
\end{equation}

\begin{figure}[!h]

\centerline{\includegraphics[width=0.7\columnwidth]{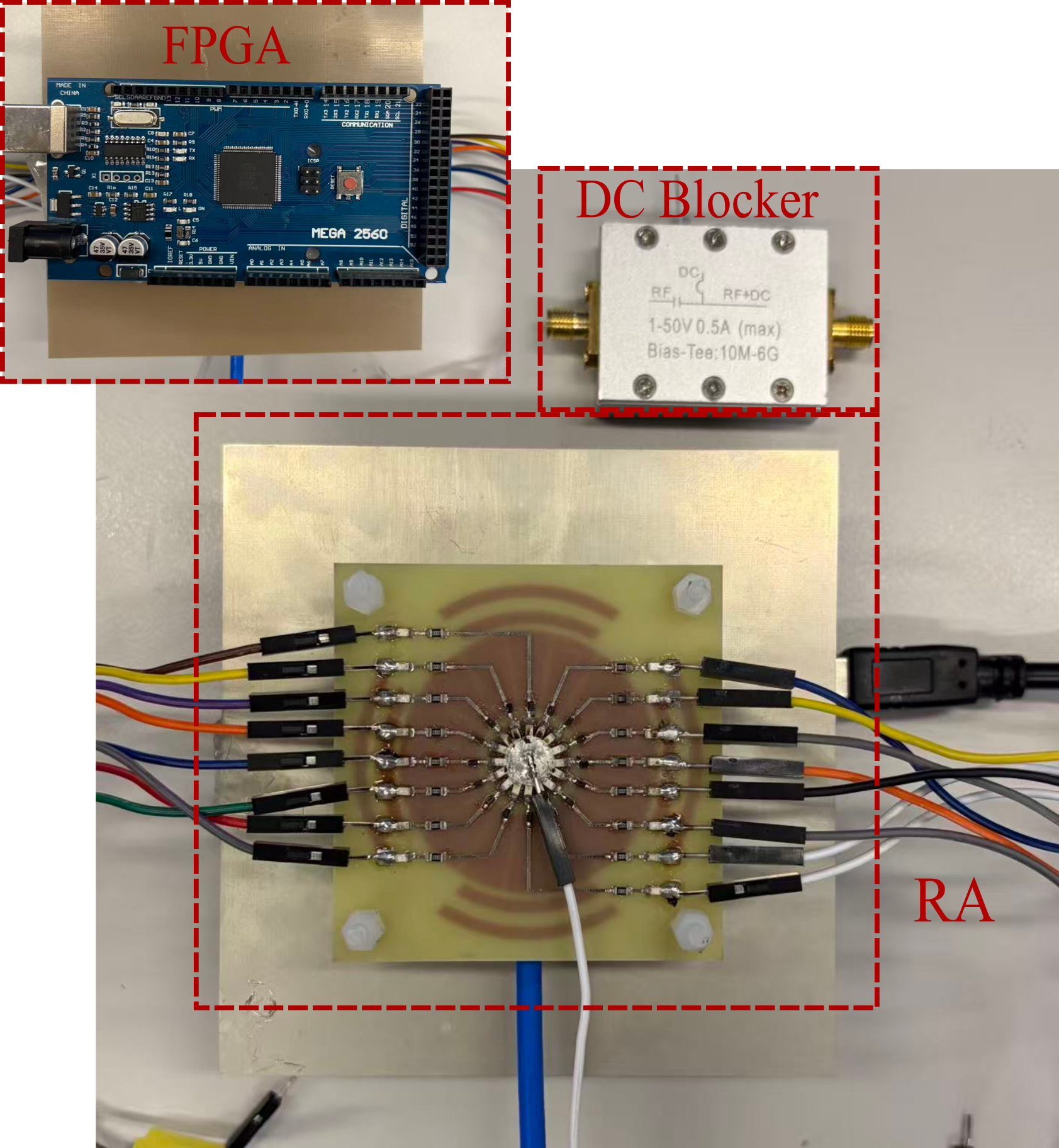}}
\caption{Fabricated prototype of the proposed antenna.}

\label{fig:prototype}
\end{figure}

The target DOAs are estimated from the $m$ dominant local maxima of $P_{\mathrm{MUSIC}}(\theta)$.  Fig.~\ref{fig:algorithm} summarizes the proposed DOA estimation procedure. Instead of using spatially separated antenna elements, the proposed antenna uses radiation pattern diversity to increase the observation dimension, while frequency diversity reduces the correlation among passive target echoes and improves the covariance rank, enabling multi-target DOA estimation with a single receive RF chain.

\section{Measurement Results}
\label{sec:measurement}
\subsection{Antenna Performance}

Fig.~\ref{fig:prototype} shows the fabricated prototype of the proposed FPRA. The DC path is controlled by an FPGA to switch the states of the PIN diodes at high speed. The RF path is connected through a DC blocker to prevent DC current from flowing into the RF measurement path. The antenna performance was simulated in Ansys HFSS 2021 and measured in an anechoic chamber. Based on the genetic algorithm optimization and measurement results, three frequencies, 2.9 GHz, 4.3 GHz, and 4.7 GHz, were selected for frequency-diversity processing. The simulated and measured $S$-parameters for different beam states are shown in Fig.~\ref{fig:s11_results}. Due to the circular symmetry of the antenna structure, Beam~1--4 and Beam~5--8 exhibit similar $S_{11}$ responses. Therefore, only the $S_{11}$ results of Beam~1--4 at the selected frequencies are presented. The simulated and measured radiation patterns at 2.9~GHz, 4.3~GHz, and 4.7~GHz are shown in Fig.~\ref{fig:radiation_measurement}. At each frequency, after optimization using a genetic algorithm, the main beam can be steered from $-40^\circ$ to $40^\circ$. The antenna gain is between 5 and 6~dBi, and $S_{11}$ remains below $-9$~dB at the corresponding operating frequencies.

\begin{figure}[!h]

\centerline{\includegraphics[width=0.7\columnwidth]{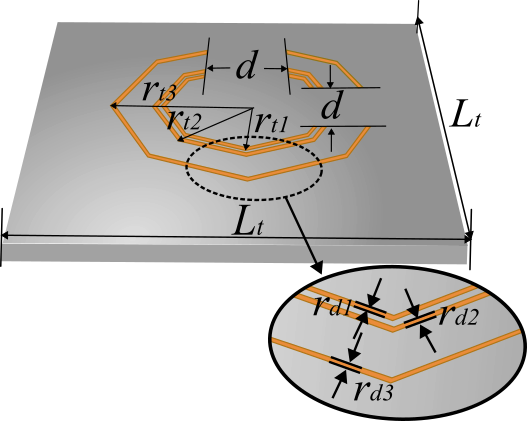}}
\caption{Schematic diagram of the RFID tag structure. The geometrical parameters are $r_{t1}=8.4~\mathrm{mm}$, $r_{t2}=5.8~\mathrm{mm}$, $r_{t3}=5.5~\mathrm{mm}$, $L_t=31.5~\mathrm{mm}$, $d=5~\mathrm{mm}$, $r_{d1}=0.1~\mathrm{mm}$, $r_{d2}=0.1~\mathrm{mm}$, and $r_{d3}=0.15~\mathrm{mm}$.}

\label{fig:rfid_structure}
\end{figure}

\begin{figure}[!h]
\centering
\begin{subfigure}[b]{0.7\columnwidth}
  \centering
  \includegraphics[width=\linewidth]{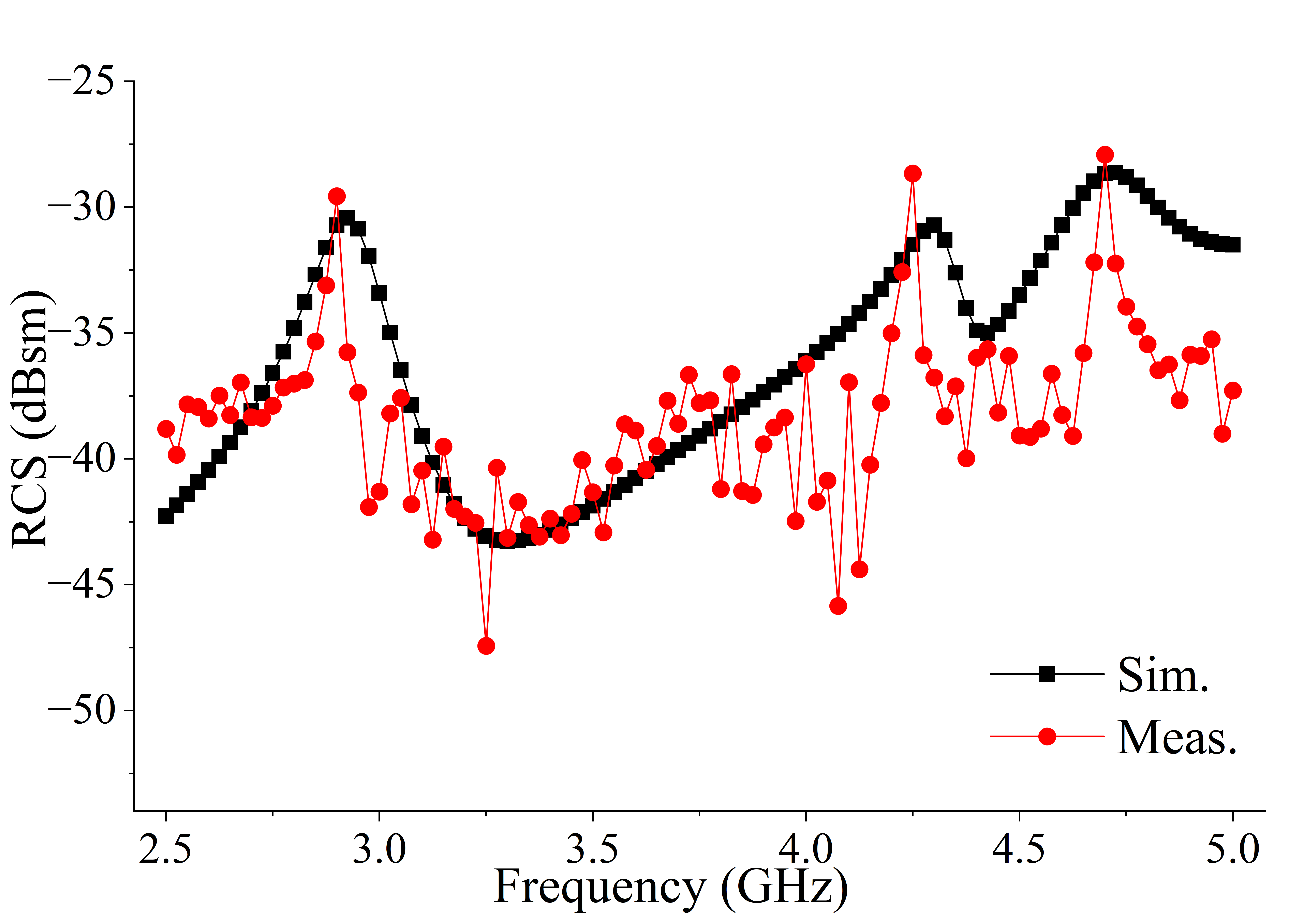}
  \caption{}
  \label{fig:rcs_vertical}
\end{subfigure}
\hfill
\begin{subfigure}[b]{0.7\columnwidth}
  \centering
  \includegraphics[width=\linewidth]{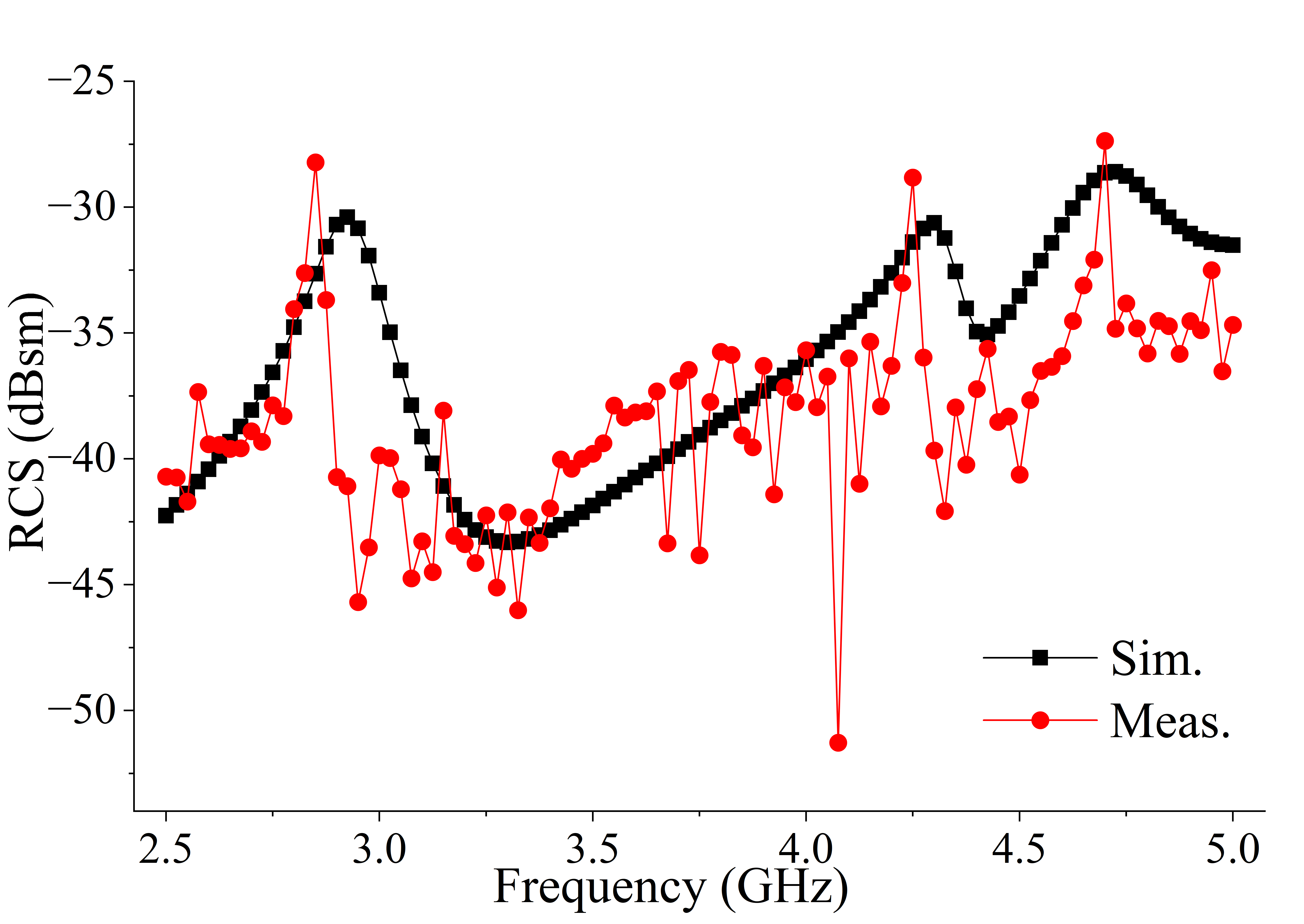}
  \caption{}
  \label{fig:rcs_horizontal}
\end{subfigure}

\caption{Measured and simulated RCS responses of the proposed chipless RFID.
(a) Vertical polarization. (b) Horizontal polarization.}
\label{fig:rcs_results}
\end{figure}

\begin{figure}[!h]

\centerline{\includegraphics[width=0.9\columnwidth]{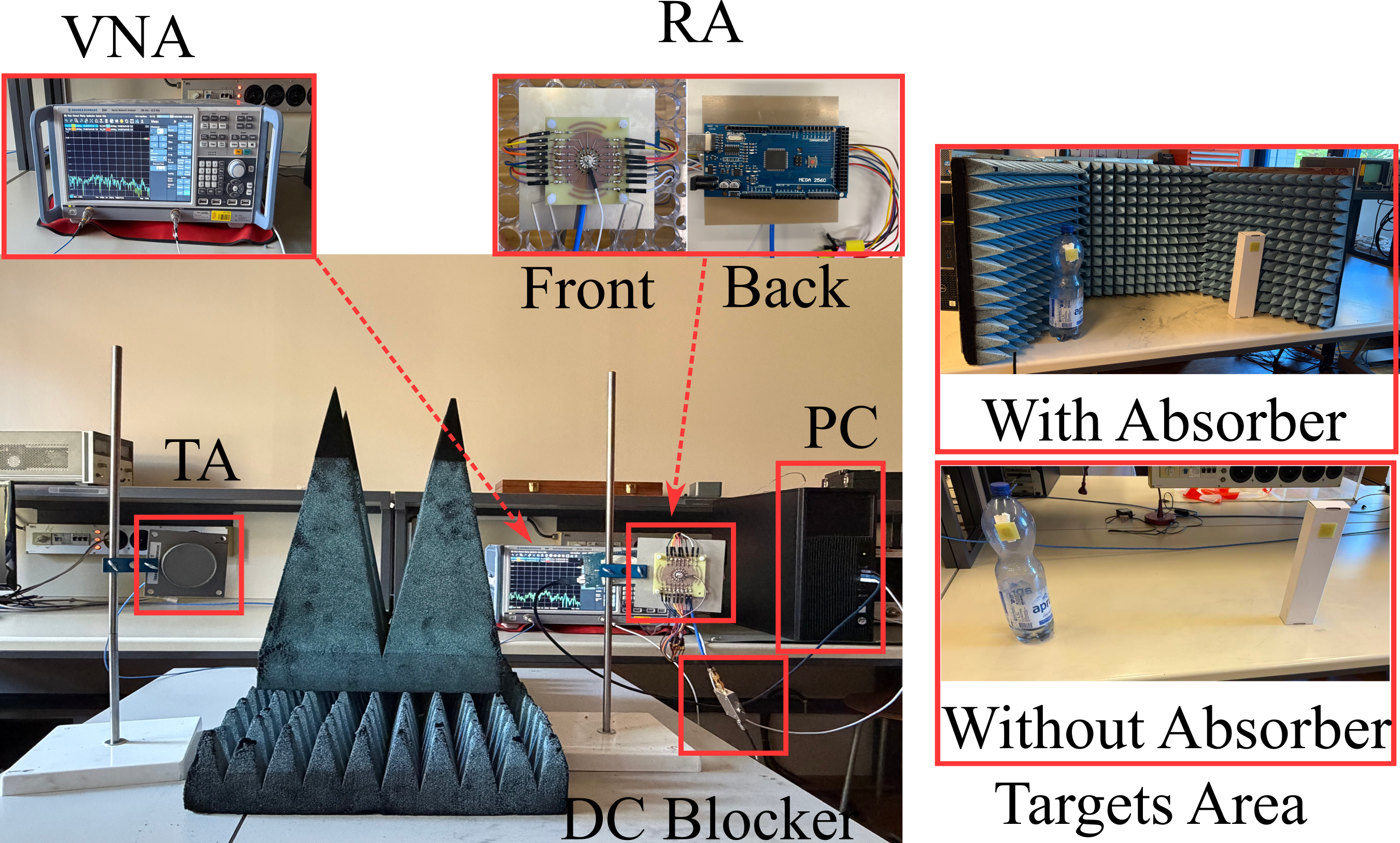}}
\caption{Experimental setup for DOA measurement.}

\label{fig:setup}
\end{figure}

\begin{figure}[!h]
\centering
\begin{subfigure}[b]{0.8\columnwidth}
  \centering
  \includegraphics[width=\linewidth]{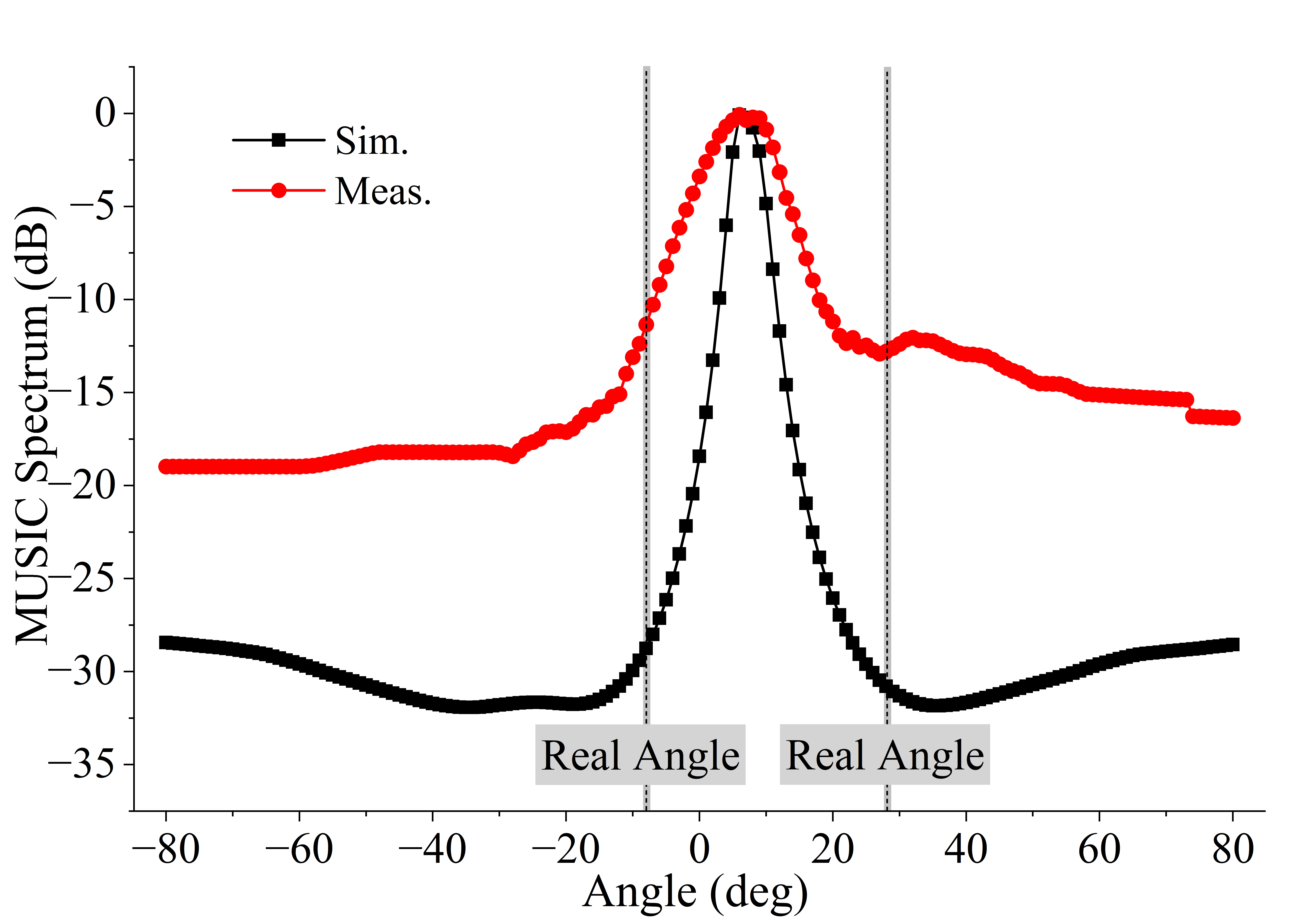}
  \caption{}
  \label{fig:doa_without_freq}
\end{subfigure}
\hfill
\begin{subfigure}[b]{0.8\columnwidth}
  \centering
  \includegraphics[width=\linewidth]{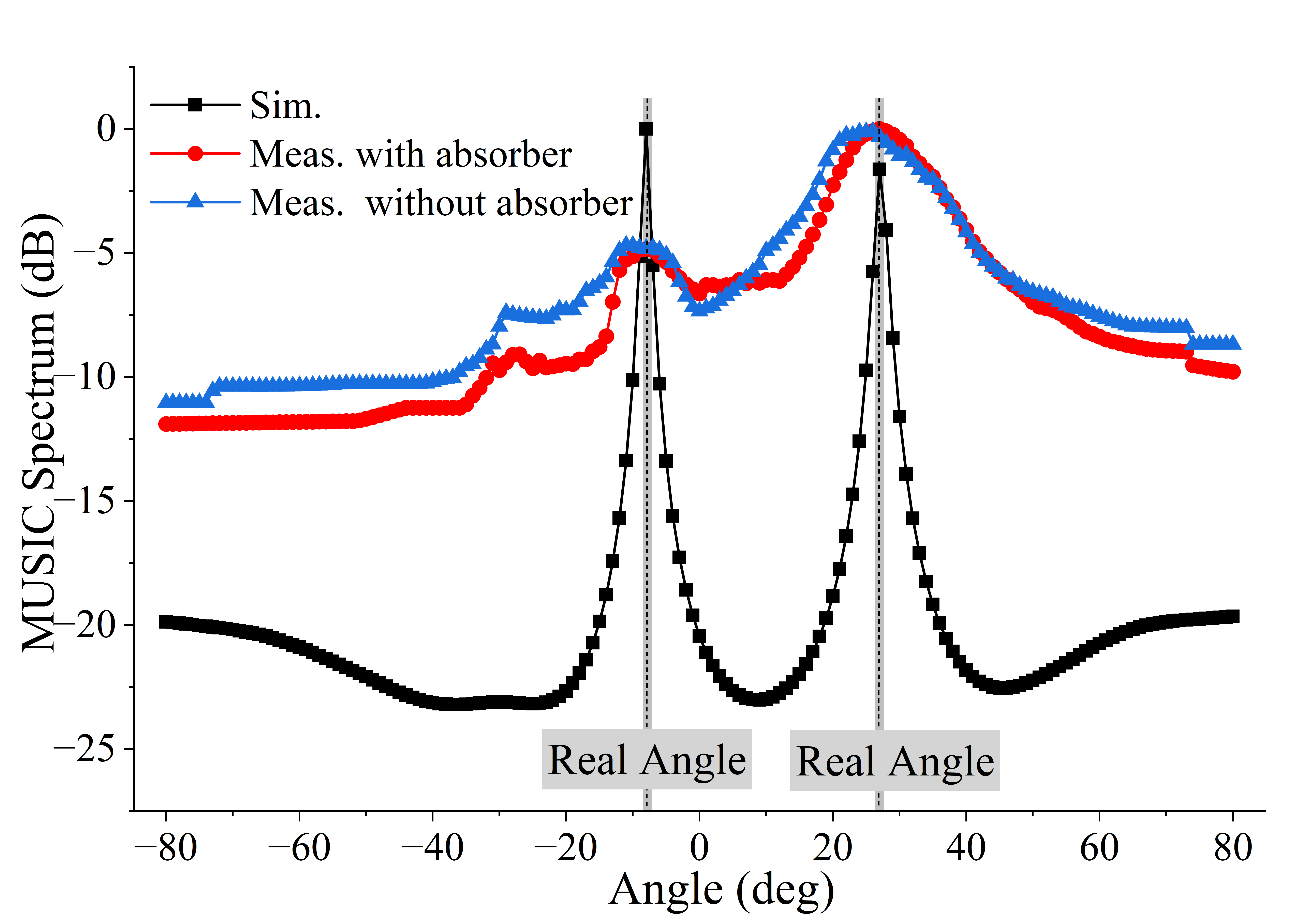}
  \caption{}
  \label{fig:doa_with_freq}
\end{subfigure}

\caption{Measured and simulated DOA estimation results obtained using the proposed single-RF-chain reconfigurable-antenna framework: (a) single-frequency processing. (b) frequency-diversity-assisted processing.}
\label{fig:doa_results}
\end{figure}

\subsection{Chipless-RFID-Based DOA Measurement Setup and Results}

To experimentally validate the proposed single-RF-chain DOA estimation framework, a chipless RFID target is designed and employed as a passive scattering object. The chipless RFID target provides a frequency-selective scattering response that can be intentionally aligned with the operating bands of the proposed FPRA, making it suitable for verifying the frequency-diversity-assisted localization capability.

The geometry of the target is shown in Fig.~\ref{fig:rfid_structure}. It consists of three regular octagonal resonators with different radii. Two orthogonal slots are etched on each octagonal resonator along the horizontal and vertical directions. Owing to this symmetric configuration, the target exhibits dual-polarized scattering characteristics. By optimizing the radii of the three resonators, distinct radar-cross-section (RCS) peaks are obtained within the operating frequency bands of the proposed FPRA. The simulated and measured RCS responses under vertical and horizontal polarizations are shown in Fig.~\ref{fig:rcs_results}.

The DOA experiment was carried out in an indoor bistatic sensing setup, as shown in  Fig.~\ref{fig:setup}. Two chipless RFID targets were placed at different azimuth angles. A commercial sinuous antenna (AEL APO-1466) was used as the transmitting antenna (TA), while the proposed FPRA was used as the receiving antenna (RA). The PIN-diode coding states of the FPRA were controlled by an FPGA connected to a PC, enabling automatic switching among different radiation-pattern and frequency states. The TA and RA were connected to the two ports of a vector network analyzer (VNA). In this bistatic configuration, the measured complex transmission coefficient \(S_{21}\) represents the received frequency-domain signal from the TA to the RA, including the amplitude and phase variations caused by the passive scattering targets. This measurement strategy is consistent with standard chipless RFID characterization methods~\cite{ref21,ref22}. At each selected frequency, the complex \(S_{21}\) samples measured under different FPRA coding states were arranged according to the coding-state index to form the observation vector \(\mathbf{y}(f_p)\), which was then processed using the DOA estimation algorithm described in Section~\ref{sec:doa_algorithm}.

The estimated DOAs of the two targets are shown in Fig.~\ref{fig:doa_results}, where the dashed vertical lines denote the true angular positions of the targets at $-8^\circ$ and $27^\circ$. Fig.~\ref{fig:doa_results}(a) shows the MUSIC spectrum obtained without frequency-diversity processing. In this case, the two correlated target echoes cannot be effectively separated, and the two targets are incorrectly identified as a single target. After jointly exploiting the selected frequency bands, as shown in Fig.~\ref{fig:doa_results}(b), two distinct target-related peaks can be observed around the true angular positions. When absorbers are placed behind the targets, the estimated DOAs are $-9^\circ$ and $28^\circ$, respectively. Without absorbers, the estimated DOAs are $-11^\circ$ and $25^\circ$, respectively. These results show that the proposed frequency-diversity-assisted single-RF-chain FPRA framework can successfully localize multiple passive chipless RFID targets in an indoor environment.

\section{Conclusion}

In this work, a frequency- and radiation-pattern-reconfigurable antenna is proposed for passive multi-target DOA estimation with a single receive RF chain. By switching the PIN diodes embedded in the circular patch, the current-concentration boundary is reconfigured, which changes both the effective resonant path and the aperture phase distribution. As a result, the proposed antenna achieves frequency tuning and beam scanning within a single compact radiator. The measured results show that the antenna can steer its main beam from $-40^\circ$ to $40^\circ$ at the selected operating frequencies of 2.9, 4.3, and 4.7 GHz, with good agreement between simulation and measurement.

Based on the reconfigurable radiation patterns, the antenna provides pattern-domain observation diversity and forms virtual spatial observations without using a conventional antenna array. Furthermore, the selected multi-frequency observations introduce phase diversity, which helps reduce the correlation among passive target echoes. Experimental results using chipless RFID targets demonstrate that the proposed framework can separate and estimate the DOAs of two correlated passive targets under a single-RF-chain architecture. Therefore, the proposed FPRA offers a compact and hardware-efficient sensing solution for future ISAC nodes requiring passive multi-target localization.

\end{document}